\newif\ifpreprint

\preprintfalse 

\ifpreprint
\documentclass[aip,jcp,amsmath,amssymb,preprint]{revtex4-1}
\else
\documentclass[aip,jcp,amsmath,amssymb,reprint]{revtex4-1}
\fi

\usepackage{xspace}
\usepackage{graphicx}
\usepackage{dcolumn}
\usepackage{bm}
\usepackage[mathlines]{lineno}
\usepackage{amsfonts}
\usepackage{pbox}
\usepackage{braket}
\usepackage{multirow}
\usepackage{threeparttable}
\usepackage{float}
\usepackage{xspace}
\usepackage{color}
\usepackage{fancyvrb}
\usepackage{adjustbox}
\usepackage{standalone}
\usepackage{tikz}
\usepackage{marvosym}
\usetikzlibrary{shapes.geometric, arrows}
\usepackage[note-name=,use-sort-key=false]{notes2bib}

\newcommand*{\cm}{cm$^{-1}$\xspace}
\newcommand*{\kcal}{kcal mol$^{-1}$\xspace}
\newcommand*{\sunit}{$E_{\rm h}^{-2}$\xspace}
\newcommand*{\Eh}{$E_{\rm h}$\xspace}
\newcommand*{\molpro}{{\scshape Molpro}\xspace}

\newcommand*{\PSI}{{\scshape Psi4}\xspace}
\newcommand*{\ambit}{{\scshape Ambit}\xspace}

\newcommand{\mref}[0]{\Psi_0}
\newcommand{\tens}[3]{{#1}_{#2}^{#3}}
\newcommand{\dfock}[1]{\epsilon_{#1}}
\newcommand{\cop}[1]{\hat{a}^\dag_{#1}}
\newcommand{\aop}[1]{\hat{a}_{#1}}
\newcommand{\sqop}[2]{\hat{a}_{#2}^{#1}}

\newcommand{\kro}[2]{\delta_{#2}^{#1}}
\newcommand{\density}[2]{\gamma_{#2}^{#1}}
\newcommand{\cdensity}[2]{\eta_{#2}^{#1}}
\newcommand{\cumulant}[2]{\lambda_{#2}^{#1}}
\newcommand{\no}[1]{ \{ {#1} \}}

\newcommand{\mkw}[0]{MK Wick's\xspace}

\newcommand{\uLDSRG}[0]{uMR-LDSRG(2)\xspace}

\makeatletter

\newbox\swb@xone
\newbox\swb@xtwo
\newbox\swb@xthree
\newbox\swb@xfour
\newdimen\swdimen@ne
\newdimen\swdimentw@

\newcommand{\acontraction}[5][1ex]{%
  \mathchoice
    {\acontraction@\displaystyle{#2}{#3}{#4}{#5}{#1}}%
    {\acontraction@\textstyle{#2}{#3}{#4}{#5}{#1}}%
    {\acontraction@\scriptstyle{#2}{#3}{#4}{#5}{#1}}%
    {\acontraction@\scriptscriptstyle{#2}{#3}{#4}{#5}{#1}}}%
\newcommand{\acontraction@}[6]{%
  \setbox\swb@xone=\hbox{${}#1{}#2{}$}%
  \setbox\swb@xtwo=\hbox{${}#1{}#3{}$}%
  \setbox\swb@xthree=\hbox{${}#1{}#4{}$}%
  \setbox\swb@xfour=\hbox{${}#1{}#5{}$}%
  \swdimen@ne=\wd\swb@xtwo%
  \advance\swdimen@ne by \wd\swb@xfour%
  \divide\swdimen@ne by 2%
  \advance\swdimen@ne by \wd\swb@xthree%
  \vbox{%
    \hbox to 0pt{%
      \kern \wd\swb@xone%
      \kern 0.5\wd\swb@xtwo%
      \acontraction@@{\swdimen@ne}{#6}%
      \hss}%
    \vskip 0.5ex
    \vskip\ht\swb@xtwo}}

\newcommand{\acontraction@@}[3][0.05em]{%
  \hbox{%
    \vrule width #1 height 0pt depth #3%
    \vrule width #2 height 0pt depth #1%
    \vrule width #1 height 0pt depth #3%
    \relax}}
\let\contraction\acontraction

\newcommand{\tcontraction}[5][1ex]{%
  \mathchoice
    {\tcontraction@\displaystyle{#2}{#3}{#4}{#5}{#1}}%
    {\tcontraction@\textstyle{#2}{#3}{#4}{#5}{#1}}%
    {\tcontraction@\scriptstyle{#2}{#3}{#4}{#5}{#1}}%
    {\tcontraction@\scriptscriptstyle{#2}{#3}{#4}{#5}{#1}}}%
\newcommand{\tcontraction@}[6]{%
  \setbox\swb@xone=\hbox{${}#1{}#2{}$}%
  \setbox\swb@xtwo=\hbox{${}#1{}#3{}$}%
  \setbox\swb@xthree=\hbox{${}#1{}#4{}$}%
  \setbox\swb@xfour=\hbox{${}#1{}#5{}$}%
  \swdimen@ne=\wd\swb@xtwo%
  \advance\swdimen@ne by \wd\swb@xfour%
  \divide\swdimen@ne by 2%
  \advance\swdimen@ne by \wd\swb@xthree%
  \vbox{%
    \hbox to 0pt{%
      \kern \wd\swb@xone%
      \kern 0.5\wd\swb@xtwo%
      \tcontraction@@{\swdimen@ne}{#6}%
      \hss}%
    \vskip 0.5ex
    \vskip\ht\swb@xtwo}}

\newcommand{\tcontraction@@}[3][0.075em]{%
  \hbox{%
    \vrule width #1 height 0pt depth #3%
    \vrule width #2 height 0pt depth #1%
    \vrule width #1 height 0pt depth #3%
    \relax}}

\makeatother

\begin{document}

\title{Towards numerically robust multireference theories: The driven similarity renormalization group truncated to one- and two-body operators}

\author{Chenyang Li}
\author{Francesco A. Evangelista}
\email{francesco.evangelista@emory.edu}
\affiliation{Department of Chemistry and Cherry Emerson Center for Scientific Computation, Emory University, Atlanta, GA, 30322}

\date{\today}

\begin{abstract}

The first nonperturbative version of the multireference driven similarity renormalization group (MR-DSRG) theory [C.~Li and F.~A.~Evangelista, J.~Chem.~Theory~Comput.~\textbf{11}, 2097 (2015)] is introduced.
The renormalization group structure of the MR-DSRG equations ensures numerical robustness and avoidance of the intruder state problem, while the connected nature of the amplitude and energy equations guarantees size consistency and extensivity.
We approximate the MR-DSRG equations by keeping only one- and two-body operators and using a linearized recursive commutator approximation of the Baker--Campbell--Hausdorff expansion [T.~Yanai and G.~K.-L.~Chan, J.~Chem.~Phys.~\textbf{124}, 194106 (2006)].
The resulting MR-LDSRG(2) equations contain only 39 terms and scales as ${\cal O}(N^2 N_{\rm P}^2 N_{\rm H}^2)$ where $N_{\rm H}$, $N_{\rm P}$, and $N$ correspond to the number of hole, particle, and total orbitals, respectively.
Benchmark MR-LDSRG(2) computations on the hydrogen fluoride and molecular nitrogen binding curves and the singlet-triplet splitting of \textit{p}-benzyne yield results comparable in accuracy to those from multireference configuration interaction,  Mukherjee multireference coupled cluster theory, and internally-contracted multireference coupled cluster theory.
\end{abstract}

\maketitle

\section{Introduction}
Striking the right balance between the theoretical treatment of static and dynamic electron correlation is a crucial requirement for predictive theories of strongly correlated electrons.\cite{Bartlett2012p182}
Consequently, the introduction of the multi-configurational self-consistent-field (MCSCF) approach\cite{Szalay2012p108} was followed by the development of a myriad of multireference (MR) theories that augment this scheme with high-level treatments of dynamic correlation.
The majority of these \textit{genuine} multireference approaches are based on the framework of effective Hamiltonian theory\cite{VanVleck:1929fp,Kemble:2005vz,Bloch:1958wc,Brandow:1967wn,Freed:1974vo,Kirtman:1981kf} and include widely adopted methods such as second-order MR perturbation theory (MRPT2)\cite{Roos1992p1218,Hirao1992p374,Davidson1994p3672,Malrieu2001p10252,Mukherjee2005p134105,Hoffmann:2009ih} and MR configuration interaction (MRCI).\cite{Knowles1988p5803,Szalay2012p108,Davidson1974p61,Gdanitz:1988et,Szalay:1993jh}
Furthermore, numerous multireference coupled cluster (MRCC) theories\cite{
Lindgren1978p33,Haque:1984dk,Monkhorst1981p1668,Piecuch2000p052506,Li:2003gi, Hubac1998p75,Pittner1999p10275,Mukherjee1998p157,Kallay2010p074103,Evangelista2006p154113,Kallay2010p074103,Hanrath2005p084102,Bartlett2007p291,Evangelista2011p114102,Hanauer2011p204111,Hoffmann2012p014108,Nooijen2011p214116,Demel:2013kz}
and alternative approaches\cite{Yanai2006p194106,Yanai2007p104107,Mazziotti2006p143002,Mazziotti2007p022505,Mazziotti2008p234103,Mazziotti2012p244} have been developed.
These approaches strive to reproduce the success of single-reference coupled cluster theory by combining a nonperturbative treatment of dynamic correlation with the requirement of size extensivity.\cite{Crawford2000p33,Bartlett2007p291}

Nevertheless, it is well appreciated that the application of multireference theories based on effective Hamiltonians presents several problems, which prevent them from being as impactful as their single-reference analogues.
The most important issue is perhaps the intruder-state problem,\cite{Malrieu1987p4930,Paldus:1993dx} which occurs when excited configurations (or determinants) become near-degenerate with the reference wave function.
In MRPT2 approaches, intruder states lead to diverging first-order excitation amplitudes and characteristic poles in potential energy surfaces.\cite{Roos1995p215,Camacho:2009cr,Camacho:2010co}
Intruders are commonly treated by shifting the energy denominators,\cite{Roos1995p215,Forsberg:1997ke} regularizing the amplitudes,\cite{Hirao2002p957,Bartlett2009p144112} modifying the zeroth-order Hamiltonian,\cite{Dyall1995p4909,Andersson:1995vv} and increasing the size of the active space.\cite{Camacho:2009cr}
However, none of these techniques have been satisfactorily generalized to the case of nonperturbative  theories (e.g., MRCC), in which intruders usually result in convergence difficulties that render these approaches inapplicable.\cite{Piecuch2000p052506,Neuscamman:2010dz}

Effective Hamiltonian theory is also affected by the problem of redundant wave function parameters.\cite{Mahapatra1998p163,Mukherjee1999p6171}
For instance, in the internally-contracted MRCC (ic-MRCC) approach,\cite{Mahapatra1998p163,Evangelista2011p114102,Hanauer2011p204111} the basis of excited configurations contains linear-dependent components which, when discarded, introduces dependencies on numerical thresholds.
A small numerical threshold induces numerical instabilities, while a large threshold may lead to discontinuous potential energy surfaces.\cite{Evangelista2011p114102}
Moreover, eliminating linearly dependent excitations requires the diagonalization of higher-order reduced density matrices, which limits the applicability of these methods to moderate numbers of active orbitals.\cite{Yanai2011p094104}
Solutions to this problem are realized only recently by either employing strongly contracted excitation operators\cite{Malrieu2001p10252,Angeli2001p297,Neuscamman:2010dz} or imposing many-body conditions.\cite{Lindgren1978p33,Bartlett1996p2652,Nooijen2011p214116}

To address the the intruder state and redundancy problems of the effective Hamiltonian formalism, we have recently begun to explore many-body theories based on the similarity renormalization group (SRG).\cite{Wilson1993p5863,Wegner2000p133,Tsukiyama:2011eo,Hergert:2016fd}
The SRG provides a systematic approach to integrate out high-energy degrees of freedom such that divergences resulting from small energy denominators are suppressed.
Inspired by the SRG, we have proposed a novel approach, the driven SRG (DSRG),\cite{Evangelista2014p054109} which combines the main features of the SRG with a computational approach closely related to coupled cluster theory.
Later, we introduced a multireference DSRG (MR-DSRG) theory that generalizes the DSRG to multiconfigurational references and investigated a second-order approximation.\cite{Li:2015iz}

The most important difference between the MR-DSRG and other multireference theories is the use of a continuous unitary transformation of the Hamiltonian controlled by an energy cutoff $\Lambda$.  This transformation excludes excitations with energy approximatively smaller than $\Lambda$, and thus, it avoids divergences caused by small denominators (intruder states).
The MR-DSRG makes also extensive use of Mukherjee and Kutzelnigg's algebra of second quantized operators that are normal ordered with respect to a multiconfigurational vacuum.\cite{Mukherjee1997p561,Mukherjee1997p432,Shamasundar2009p174109,Mukherjee2010p234107,Mukherjee2013p62,Kutzelnigg:2010iu}
Building upon this algebra, the MR-DSRG equations are formulated in Fock space\cite{Kutzelnigg:1982kr,*Kutzelnigg:1983dr,*Kutzelnigg:1984eg,*Kutzelnigg:1985fj,Stolarczyk:1985fk,*Stolarczyk:1985ct,*Stolarczyk:1988ci,*Stolarczyk:1988cv} as a set of many-body conditions.\cite{Lindgren1978p33,Bartlett1996p2652,Nooijen2011p214116}
The use of many-body conditions leads to an equal number of equations and unknowns, and therefore, it guarantees that the MR-DSRG is free from the redundancy problem.


Our initial work on the MR-DSRG examined the accuracy and numerical robustness of a second-order approximation.
The goal of this work is to go beyond a perturbative treatment of dynamic electron correlation and explore one of simplest MR-DSRG nonperturbative schemes.
The resulting model---designated as MR-LDSRG(2)---retains all of the one- and two-body components of the renormalized Hamiltonian and expands the MR-DSRG transformation in terms of a linear recursive commutator approximation.\cite{Yanai2006p194106,Evangelista:2012hz}
The MR-LDSRG(2) energy may be evaluated with a computational procedure that has a computational scaling analogous to that of the coupled cluster approach with singles and doubles (CCSD).
In addition, the MR-LDSRG(2) approach requires only the knowledge of the one-particle density matrix and the two- and three-body density cumulants\cite{Mukherjee1997p432,Kutzelnigg:2010iu,Hanauer2012p50} of the reference wave function.

We start from an overview of the MR-DSRG formulation and introduce the MR-LDSRG(2) model in Sec.~\ref{sec:theory}.
Section~\ref{sec:implementation} presents our pilot implementation and discusses the scaling of the MR-LDSRG(2) approach.
Applications of the MR-LDSRG(2) to the singlet ground-state potential energy curves of HF and N$_2$, and the singlet-triplet splitting of \textit{p}-benzyne are reported in Sec.~\ref{sec:results}, where computational details are given in Sec.~\ref{sec:comput_detail}.
Finally in Secs.~\ref{sec:comparison} and \ref{sec:conclusions}, we compare the MR-DSRG ans{\"a}tz to other methods based on internally contracted formalism, and discuss some future developments of the MR-DSRG theory.

\section{Theory}
\label{sec:theory}

\subsection{Basic notation}
\label{sec:orbitalspaces}

We define the Fermi vacuum as a multideterminantal wave function $\ket{\mref}$ with respect to which all second quantized operators are normal ordered:
\begin{align}
\label{eq:ref}
    \ket{\mref} = \sum_{\mu = 1}^{d} c_{\mu} \ket{\Phi^{\mu}}.
\end{align}
In Eq.~\eqref{eq:ref}, the set of determinants $\{\Phi^{\mu}\}$ form a complete active space (CAS).
The orbital space $\{\phi^p, p =1, \ldots, N\}$ is thus partitioned into three subsets: core ($\bf C$), active ($\bf A$), and virtual ($\bf V$).
For convenience, we also define two composite spaces: hole ($\bf H = C \cup A$) and particle ($\bf P = A \cup V$).
The orbital indices corresponding to these spaces are listed in Table \ref{tab:orbital_space}.

The bare Hamiltonian normal ordered with respect to $\mref$ is given by:
\begin{align}
\label{eq:bareH}
    \hat{H} &= E_{0} + \sum_{pq} \tens{f}{p}{q} \no{\sqop{p}{q}} + \frac{1}{4} \sum_{pqrs} \tens{v}{pq}{rs} \no{\sqop{pq}{rs}},
\end{align} 
where $E_{0} = \bra{\mref} \hat{H} \ket{\mref}$ is the reference energy and $\{ \sqop{ab\cdots}{ij\cdots} \} = \{ \cop{a} \cop{b} \cdots \aop{j} \aop{i} \}$ stands for a string of normal-ordered creation ($\hat{a}^\dagger$) and annihilation ($\hat{a}$) operators.
In Eq.~\eqref{eq:bareH}, we have introduced the matrix element of the generalized Fock matrix $\tens{f}{p}{q}$:
\begin{align}
\label{eq:fock}
    \tens{f}{p}{q} = \tens{h}{p}{q} + \sum_{rs} \tens{v}{pr}{qs} \density{r}{s},
\end{align}
where $\density{p}{q} = \braket{\mref| \sqop{p}{q} |\mref}$, $\tens{h}{p}{q} = \braket{\phi_{p}|\hat{h}|\phi_q}$, and $\tens{v}{pq}{rs}= \braket{\phi_{p}\phi_{q}\|\phi_{r}\phi_{s}}$ are respectively the one-particle density matrix element of the reference, the one-electron integrals, and the antisymmetrized two-electron integrals.
For convenience, we also assume to work with a semicanonical orbital basis such that the core, active, and virtual blocks of the generalized Fock matrix are diagonal.

\begin{table}[ht!]
\centering
\setlength{\tabcolsep}{2pt}
\renewcommand{\arraystretch}{1.25}
\caption{Definition of the orbital spaces employed in this work.}
\label{tab:orbital_space}
\begin{tabular*}{\columnwidth}{@{\extracolsep{\stretch{1}}}*{1}{l}*{4}{c}@{}}
\hline

\hline
Space & Symbol & Dimension & Indices & Description \\
\hline
Core & $\bf C$ & $N_{\rm C}$ & $m, n$ & Doubly occupied \\
 Active & $\bf A$ & $N_{\rm A}$ & $u, v, w, x, y, z$ & Partially occupied \\
 Virtual & $\bf V$ & $N_{\rm V}$ & $e, f$ & Unoccupied \\
 Hole & $\bf H$ & $N_{\rm H}$ & $i, j, k, l$ & $\bf H = C \cup A$ \\
 Particle & $\bf P$ & $N_{\rm P}$ & $a, b, c, d$ & $\bf P = A \cup V$ \\
 General & $\bf G$ & $N$ & $p, q, r, s$ & $\bf G = H \cup V$ \\
\hline

\hline
\end{tabular*}
\end{table}

\subsection{MR-DSRG Theory}
\label{sec:mrdsrg}

In the unitary MR-DSRG ansatz,\cite{Evangelista2014p054109,Li:2015iz} the bare Hamiltonian $(\hat{H})$ is partially block-diagonalized by a unitary transformation.  The unitary operator that performs this transformation is written in an exponential form, $e^{\hat{A}(s)}$, where $\hat{A}(s)$ is a $s$-dependent anti-Hermitian operator.
The flow variable $s$ is defined in the range [0,$\infty$) and controls the extent of the DSRG transformation.
The DSRG unitary transformation yields an effective (or renormalized) Hamiltonian $\bar{H}(s)$ (see Refs. \citenum{Evangelista2014p054109} and \citenum{Li:2015iz} for details), which may be partitioned into a sum of diagonal $\bar{H}^{\rm D} (s)$ and non-diagonal $\bar{H}^{\rm N} (s)$ components:\cite{Kutzelnigg2010p299,Kutzelnigg2009p3858}
\begin{align}
    \bar{H}(s) = e^{-\hat{A} (s)} \hat{H} e^{\hat{A} (s)} = \bar{H}^{\rm D} (s) + \bar{H}^{\rm N} (s). \label{eq:H2Hbar}
\end{align}
The diagonal component contains only the pure excitation and de-excitation diagrams and couples the reference $\ket{\mref}$ to excited configurations of the form $\{\sqop{ab\cdots}{ij\cdots}\}\ket{\mref}$.\cite{Evangelista2014p054109,Li:2015iz}

The MR-DSRG transformation [Eq.~\eqref{eq:H2Hbar}] is determined by the flow equation:
\begin{align}
\label{eq:dsrg}
    \bar{H}^{\rm N}(s) = \hat{R}(s),
\end{align}
where $\hat{R}(s)$ is the so-called \textit{source operator}, a Hermitian operator that drives the off-diagonal components of $\bar{H}(s)$ to zero, that is $\lim_{s \rightarrow \infty} \bar{H}^{\rm N}(s) = 0$.
The source operator $\hat{R}(s)$ is required to perform a renormalization transformation, that is, to decouple only those excited configurations that differ from the reference by an energy larger than the cutoff $\Lambda = s^{-1/2}$.\cite{Evangelista2014p054109,Kehrein2010book}
These two requirements do not identify a unique form for $\hat{R}(s)$.  Therefore, in our work we use a source operator designed to reproduce some of the features of the SRG approach (see below).\cite{Evangelista2014p054109}
Once $\hat{R}(s)$ is specified, the MR-DSRG equation implicitly determines the anti-Hermitian operator $\hat{A}(s)$ and the renormalized Hamiltonian [Eq.~\eqref{eq:H2Hbar}].
As we shall discuss more in detail in Sec.~\ref{sec:LDSRG}, the DSRG equation should be understood as a collection of many-body conditions,\cite{Lindgren1978p33,Bartlett1996p2652,Nooijen2011p214116} where the coefficients associated to the same normal-ordered second-quantized operators on the left and right side of Eq.~\eqref{eq:dsrg} are set equal to each other. 

The electronic energy for a given reference $\mref$ is computed as the expectation value of the DSRG transformed Hamiltonian $\bar{H}(s)$:
\begin{align}
\label{eq:energy}
    E(s) = \braket{\mref| \bar{H}(s) |\mref}.
\end{align}
The \textit{relaxed} MR-DSRG energy is obtained using coefficients that diagonalize $\bar{H}(s)$ within the space of reference determinants:
\begin{align}
\label{eq:eigen}
    \sum_{\mu}^{d} \braket{\Phi_{\nu} | \bar{H}(s) | \Phi^{\mu}} c_{\mu} = E(s) c_{\nu}.
\end{align}
Note that computing the relaxed MR-DSRG energy requires the simultaneous solution of the MR-DSRG equation [Eq.~\eqref{eq:dsrg}] and the energy eigenvalue equation [Eq.~\eqref{eq:eigen}].
In addition, we also consider the \textit{unrelaxed} energy, which is obtained by evaluating $E(s)$ using reference coefficients from a CAS configuration interaction (CASCI) or CAS self-consistent field (CASSCF)\cite{Roos1980p157} computation.
Results from unrelaxed computations will be denoted by the prefix ``u'' (for example, uMR-DSRG).

\subsection{The linearized MR-DSRG scheme with one- and two-body operators [MR-LDSRG(2)]}
\label{sec:LDSRG}

The essence of the MR-DSRG framework is to solve the DSRG equation [Eq.~\eqref{eq:dsrg}] using a many-body formalism.\cite{Lindgren1978p33,Bartlett1996p2652,Nooijen2011p214116,Evangelista2014p054109}
As in the case of configuration interaction and coupled cluster theory, the MR-DSRG equations can be systematically truncated  to form a hierarchy of increasingly accurate methods [MR-DSRG($n$), $n = 2, 3, \ldots$].
To this end, the anti-Hermitian operator $\hat{A}(s)$ is written in terms of a cluster operator [$\hat{T}(s)$] as:
\begin{align}
\label{eq:A}
    \hat{A}(s) &= \hat{T}(s) - \hat{T}^\dag (s),
\end{align}
and the cluster operator $\hat{T} (s)$ is a sum of excitation operators up to rank $n$:
\begin{equation}
    \hat{T} (s) = \sum_{k=1}^{n} \hat{T}_k(s), \label{eq:T_mb}
\end{equation}
where each $k$-fold component [$\hat{T}_k(s)$] is defined as:
\begin{align}
\label{eq:T_k}
    \hat{T}_k(s) = \frac{1}{(k!)^2} \sum_{ij\cdots}^{\mathbf{H}} \sum_{ab\cdots}^{\mathbf{P}} \tens{t}{ab\cdots}{ij\cdots} (s) \no{\sqop{ab\cdots}{ij\cdots}}.
\end{align}
As shown in Eq.~\eqref{eq:T_k}, $\hat{T}_k(s)$ incorporates strings of $k$ normal-ordered creation and annihilation operators ($\no{\sqop{ab\cdots}{ij\cdots}}$), and each operator associates to a tensor [$\tens{t}{ab\cdots}{ij\cdots} (s)$] that is antisymmetric with respect to distinct permutations of upper and lower indices.
\textit{Internal} cluster amplitudes that are labeled only by active orbital indices are redundant since they only change the reference coefficients.  Therefore, internal amplitudes are set to zero, that is $\tens{t}{uv\cdots}{xy\cdots} (s) = 0$ for $uv \cdots, xy\cdots \in \mathbf{A}$.

The left-hand-side of the DSRG equation [Eq.~\eqref{eq:dsrg}] contains the DSRG Hamiltonian $\bar{H}(s)$, which may be expressed as a series of commutators of $\hat{H}$ and $\hat{A}(s)$ using the Baker--Campbell--Hausdorff (BCH) formula:
\begin{align}
\label{eq:exactBCH}
    \bar{H}(s) = \hat{H} + \sum_{k=1}^{\infty} \frac{1}{k!} \underbrace{[\cdots[[\hat{H}, \hat{A}(s)], \hat{A}(s)], \cdots]}_{k\text{ nested commutators}}.
\end{align}
The DSRG Hamiltonian is a general Hermitian many-body operator and may be expressed in terms of normal-ordered components of different rank:\cite{Nooijen2011p214116,Datta2012p204107,Demel:2013kz}
\begin{equation}
\label{eq:Hbar_mb}
\bar{H}(s) = E(s) + \sum_{k=1}^{N} \bar{H}_k(s).
\end{equation}
In Eq.~\eqref{eq:Hbar_mb} the term $\bar{H}_k(s)$ collects all the $k$-body components of $\bar{H}(s)$:
\begin{equation}
\bar{H}_k(s) =  \frac{1}{(k!)^2} \sum_{pqrs\cdots}^{\mathbf{G}} \tens{\bar{H}}{pq\cdots}{rs\cdots} (s) \no{\sqop{pq\cdots}{rs\cdots}}
\end{equation}

The source operator that appears on the right-hand-side of Eq.~\eqref{eq:dsrg} may be expanded in a similar way,
\begin{align}
    \hat{R}(s) &= \sum_{k=1}^{N} \hat{R}_k(s), \label{eq:R_mb}\\
    \hat{R}_k(s) &= \frac{1}{(k!)^2} \sum_{ij\cdots}^{\bf H} \sum_{ab\cdots}^{\bf P} \tens{r}{ab\cdots}{ij\cdots} (s) (\no{\sqop{ab\cdots}{ij\cdots}} + \no{\sqop{ij\cdots}{ab\cdots}}), \label{eq:R_k}
\end{align}
where the coefficients $\tens{r}{ab\cdots}{ij\cdots} (s)$ are given by:
\begin{align}
\label{eq:source_tensor}
    \tens{r}{ab\cdots}{ij\cdots}(s) &= [\tens{\bar{H}}{ab\cdots}{ij\cdots}(s) + \tens{t}{ab\cdots}{ij\cdots}(s) \tens{\Delta}{ab\cdots}{ij\cdots}] e^{-s (\tens{\Delta}{ab\cdots}{ij\cdots})^2},
\end{align}
where $\tens{\Delta}{ab\cdots}{ij\cdots} = \epsilon_i + \epsilon_j + \cdots - \epsilon_a- \epsilon_b - \cdots$ is a generalized M{\o}ller--Plesset denominator and $\epsilon_p = \tens{f}{p}{p}$ is the energy of orbital $\phi^p$.
The source operator defined by Eq.~\eqref{eq:source_tensor} reproduces the unitary transformation achieved by the single-reference SRG expanded to second order.\cite{Wilson1993p5863,Wegner2000p133,Tsukiyama:2011eo,Evangelista2014p054109}
It is important to note that the equation for the source operator given in Eq.~\eqref{eq:source_tensor} is valid only in the semicanonical basis.\cite{Handy1989p185,Li:2015iz}
As discussed in Appendix~\ref{appendix:orbital_rotation}, with some extra effort it is possible to formulate an orbital invariant version of the MR-DSRG theory that allows to use natural or other types of noncanonical orbitals.

After inserting the Eqs.~\eqref{eq:Hbar_mb}--\eqref{eq:source_tensor} into the DSRG equation [Eq.~\eqref{eq:dsrg}], we obtain the following set of many-body conditions:
\begin{equation}
\label{eq:many-body}
    \tens{\bar{H}}{ab\cdots}{ij\cdots} (s) = \tens{r}{ab\cdots}{ij\cdots} (s), \text{ for } ij\cdots \in \mathbf{H}, ab\cdots \in \mathbf{P}.
\end{equation}

In this work we consider the MR-DSRG truncated to one- and two-body operators, that is, we approximate the cluster operator as $\hat{T} \approx \hat{T}_1 + \hat{T}_2$.
Consequently, the DSRG equations reduce to $\tens{\bar{H}}{a}{i} (s) = \tens{r}{a}{i} (s)$ and $\tens{\bar{H}}{ab}{ij} (s) = \tens{r}{ab}{ij} (s)$.
At the same time, to produce a computationally viable method it is also necessary to truncate the BCH expansion of $\bar{H}(s)$.
Since the operator $\hat{A}(s)$ contains both excitation and de-excitation operators ($\hat{T}$ and $\hat{T}^\dagger$), the BCH expansion of the DSRG Hamiltonian does not terminate, thus, making the exact evaluation of $\bar{H}(s)$ impractical.
This issue also arises in unitary versions of single- and multireference coupled cluster theories.\cite{Bartlett:1989iz,Taube:2006bi,Yanai2006p194106,Yanai2007p104107,Hoffmann2012p014108}
Following the approach of Yanai and Chan,\cite{Yanai2006p194106,Yanai2007p104107} we approximate each commutator that enters into the BCH formula with its one- and two-body components (indicated with the subscript ``1,2''):
\begin{align}
\label{eq:BCH}
    \bar{H}(s)_{1,2} = \hat{H} + \sum_{k=1}^{\infty} \frac{1}{k!} \underbrace{[\cdots[[\hat{H}, \hat{A}(s)]_{1,2}, \hat{A}(s)]_{1,2}, \cdots]_{1,2}}_{k\text{ nested commutators}}.
\end{align}
This recursive approximation is consistent with the level of truncation of the cluster operator and leads to a practical and efficient computational scheme.
We name this truncated MR-DSRG approach as MR-LDSRG(2) where the ``L'' indicates the linear commutator approximation and ``(2)'' denotes that the DSRG equations are truncated to one- and two-body operators.

\subsection{Structure of the MR-DSRG equations}
\label{sec:structure}

In this section we compare the structure of the MR-DSRG equations to those of the single-reference coupled cluster (CC) theory.
To evaluate the commutators in the DSRG Hamiltonian [Eq.~\eqref{eq:BCH}], we use the Mukherjee--Kutzelnigg generalized Wick's theorem (\mkw theorem).\cite{Mukherjee1997p432,Mukherjee2010p234107}
For two normal-ordered second-quantized operators (e.g., $\no{\hat{X}}$ and $\no{\hat{Y}}$), the \mkw theorem allows us to express the product $\no{\hat{X}} \no{\hat{Y}}$ as the normal-ordered product $\no{\hat{X}\hat{Y}}$ plus a sum over contractions of normal ordered operators:
\begin{equation}
\label{eq:wick}
\begin{split}
    \no{\hat{X}} \no{\hat{Y}} =& \no{\hat{X}\hat{Y}}
    + \sum_{\text{single} \atop \text{pairs}} \no{ 
        \contraction[0.8ex]{}{\hat{X}}{\,\,}{\hat{Y}}
        \hat{X}\,\,\hat{Y}} 
    + \sum_{\text{double} \atop \text{pairs}} \no{
        \contraction[0.8ex]{\!}{\hat{X}}{\,\,}{\hat{Y}}
        \contraction[1.5ex]{\,}{\hat{X}}{\,\,}{\hat{Y}}
        \hat{X}\,\,\hat{Y}}\\
    &+ \sum_{\text{single} \atop \text{4-leg}} \no{
        \contraction[0.8ex]{\!}{\hat{X}}{\,\,}{\hat{Y}}
        \contraction[0.8ex]{\,}{\hat{X}}{\,\,}{\hat{Y}}
        \hat{X}\,\,\hat{Y}}
        + \sum_{\text{single} \atop \text{pairs}}  \sum_{\text{single} \atop \text{4-leg}} \no{
        \contraction[0.8ex]{\!}{\hat{X}}{\,\,}{\hat{Y}}
        \contraction[0.8ex]{\,}{\hat{X}}{\,\,}{\hat{Y}}
        \contraction[1.5ex]{\,\,\,}{\hat{X}}{\,\,}{\hat{Y}}
        \,\hat{X}\,\,\hat{Y}}
    + \cdots .
\end{split}
\end{equation}
When compared to the traditional Wick's theorem used in single-reference (SR) theories,\cite{Crawford2000p33} the \mkw theorem contains two new aspects.
Firstly, contrary to the single-reference case in which pairwise contractions introduce a Kronecker delta  ($\kro{}{}$), in the multireference case pairwise contractions give either a one-particle ($\boldsymbol{\gamma}_1$) or one-hole ($\boldsymbol{\eta}_1$) density matrix:
\begin{align}
    \contraction{}{\hat{a}}{{}^{p}}{\hat{a}}
    \cop{p} \aop{q}
    &= \density{p}{q},\\
    \contraction{}{\hat{a}}{{}_{q}}{\hat{a}}
    \aop{q} \cop{p}
    &= \cdensity{p}{q} = \kro{p}{q} - \density{p}{q}.
\end{align}
Secondly, new multi-legged contractions appear, each of which contains $2k$-legs ($k\geq2$) and pairs $k$ creation operators with $k$ annihilation operators.  These new contractions correspond to elements of the $k$-body density cumulant ($\boldsymbol{\lambda}_k$) of the reference $\mref$.
It is important to note that cumulant contractions span only those orbitals that are partially occupied in the reference.  Hence, for a complete active space reference, cumulant contractions only connect operators labeled by active indices.

It is instructive and insightful to compare the structure of the MR-DSRG Hamiltonian obtained with the \mkw theorem with the similarity transformed Hamiltonian of CC theory.
In the MR-DSRG, each commutator in the BCH expansion contain contributions of the form
\begin{equation}
\label{eq:commutator_H_A}
[\hat{H},\hat{A}] = [\hat{H},\hat{T} - \hat{T}^\dagger]
= \hat{H}\hat{T} - \hat{T}\hat{H} - \hat{H}\hat{T}^\dagger + \hat{T}^\dagger \hat{H}.
\end{equation}
One may identify two classes of terms that arise from the application of the \mkw theorem to each product  of operators that appear in Eq.~\eqref{eq:commutator_H_A}.
The first class contains only pairwise contractions.
These terms have the same structure of the CC contributions, except for the fact that their expressions contain matrix elements of $\boldsymbol{\gamma}_1$ and $\boldsymbol{\eta}_1$.
However, by an appropriate redefinition of the cluster amplitudes, these terms are equivalent to the single-reference coupled cluster equations.
\bibnote{
For example, in the MR-DSRG the expectation value of $[\hat{F}, \hat{T}_1]$ is given by:
\begin{equation*}
\bra{\mref}[\hat{F}, \hat{T}_1]\ket{\mref} = \sum_{ij}^{\mathbf{H}} \sum_{ab}^{\mathbf{P}}\, \tens{f}{j}{b} \tens{t}{a}{i} \density{j}{i}\cdensity{a}{b}, \nonumber
\end{equation*}
where $\hat{F}$ is the generalized Fock operator.
If we define dressed singles amplitudes as:
$\tens{\tilde{t}}{b}{j} = \sum_{i}^{\mathbf{H}} \sum_{a}^{\mathbf{P}}\, \tens{t}{a}{i} \density{j}{i}\cdensity{a}{b}$, then the above equation may be written as 
$\bra{\mref}[\hat{F}, \hat{T}_1]\ket{\mref} = \sum_{i}^{\mathbf{H}} \sum_{a}^{\mathbf{P}}\, \tens{f}{i}{a} \tens{\tilde{t}}{a}{i}$, 
which has the same form of the single-reference coupled cluster contribution to the energy:
\begin{equation*}
\bra{\Phi}[\hat{F}, \hat{T}_1] \ket{\Phi} = \sum_{i}^{\mathbf{O}} \sum_{a}^{\mathbf{V}}\, \tens{f}{i}{a} \tens{t}{a}{i},
\end{equation*}
where $\mathbf{O}$ and $\mathbf{V}$ are respectively the set of occupied and virtual orbitals for Slater determinant $\Phi$.

It is also possible to show that for a complete or incomplete active space, the MR-DSRG equations contain all the contributions that appear in CC theory.
Taking advantage of the structure of the one-particle and one-hole density matrices, each sum over pairwise contractions can be split into contractions over contractions over core, active, and virtual orbitals.  For example:
\begin{equation*}
 \sum_{\text{single} \atop \text{pairs}} \no{ 
        \contraction[0.8ex]{}{\hat{X}}{\,}{\hat{Y}}
        \hat{X}\,\hat{Y}}
=
 \sum_{\text{single} \atop \text{pairs}}^{\mathbf{C}} \no{ 
        \contraction[0.8ex]{}{\hat{X}}{\,}{\hat{Y}}
        \hat{X}\,\hat{Y}}
+
 \sum_{\text{single} \atop \text{pairs}}^\mathbf{A} \no{ 
        \contraction[0.8ex]{}{\hat{X}}{\,}{\hat{Y}}
        \hat{X}\,\hat{Y}}
+ \sum_{\text{single} \atop \text{pairs}}^{\mathbf{V}} \no{ 
        \contraction[0.8ex]{}{\hat{X}}{\,}{\hat{Y}}
        \hat{X}\,\hat{Y}}        .
\end{equation*}
Since for core orbitals $\density{m}{n} = \kro{m}{n}$ and $\cdensity{m}{n} = 0$, while for virtual orbitals $\density{e}{f} = 0$ and $\cdensity{e}{f} = \kro{e}{f}$, pairwise contractions of core and virtual orbitals follow the same rules of the traditional Wick's theorem.
Thus, contractions of commutators of $\hat{H}$ with $\hat{T}$ that involve only core and virtual orbitals will yield terms that are equivalent to those that appear in single-reference coupled cluster theory.
}

The second class of terms that arises from \mkw theorem [Eq.~\eqref{eq:wick}] consists of contractions that involve cumulants.
These contractions are not contained in the single-reference CC equations, and they increase the algebraic complexity of multireference internally-contracted approaches.
Nevertheless, for CAS-CI and CASSCF references cumulants can only contract second-quantized operators labeled by active indices, which implies that the computational cost of these additional terms is proportional to a polynomial in the number of active orbitals.

Another point of divergence between the MR-DSRG and CC equations arises from the mixed particle-hole character of the operators labeled by active orbital indices ($\aop{u}$ and $\cop{v}$) that enter in the definition of the cluster operator.
These operators do not fall in the traditional categories of vacuum creation and annihilation operators because, in general, they neither create nor annihilate the reference $\mref$.
Consequently, commutators of the form $[\no{\hat{X}}, \no{\hat{T}}]$ cannot be simply expressed as the connected part of  $\no{\hat{X}}\no{\hat{T}}$, like in the coupled cluster theory.
Instead, one must also consider the connected contribution from the product $\no{\hat{T}}\no{\hat{X}}$:
\begin{align}
    [\no{\hat{X}}, \no{\hat{T}}] = (\no{\hat{X}}\no{\hat{T}} - \no{\hat{T}}\no{\hat{X}})_{\rm connected}.
\end{align}
In the evaluation of commutators of the form $[\no{\hat{X}}, \no{\hat{T}}]$ several simplifications may apply.
For example, single pairwise contractions give a Kronecker delta, while single multi-leg contractions give null contributions.

\section{Implementation}
\label{sec:implementation}

The MR-LDSRG(2) method is implemented as a \PSI\cite{PSI4} plugin augmented with the open-source tensor library \ambit.\cite{AMBIT2015} 
\begin{figure}[h]
\centering
\includegraphics[width=0.90\columnwidth]{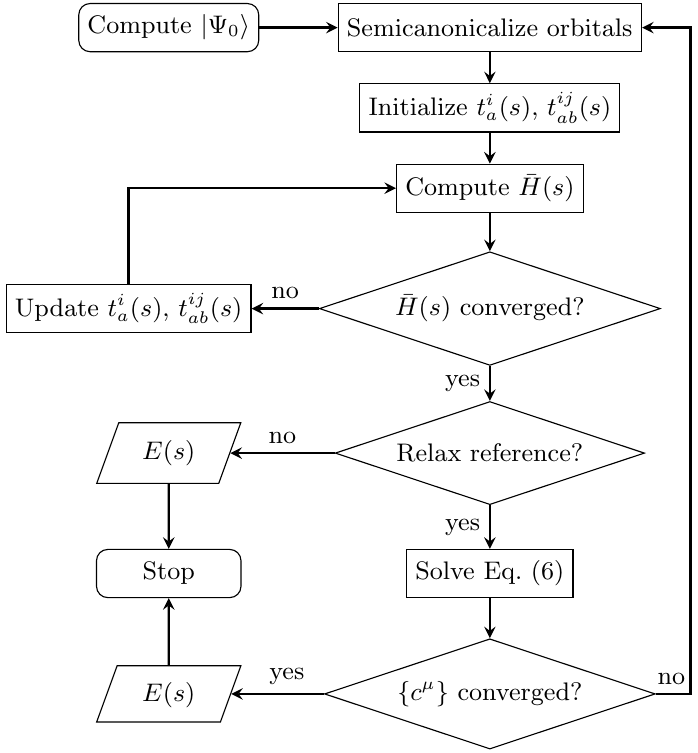}
\caption{The algorithm used to evaluate the MR-LDSRG(2) energy.}
\label{fig:flowchart}
\end{figure}
The MR-LDSRG(2) energy and cluster amplitudes are computed via an iterative procedure briefly summarized in Fig.~\ref{fig:flowchart}.
The first step is determining the reference wave function $\ket{\mref}$ in the semicanonical basis, and computing the one-particle density matrix, and two- and three-body density cumulants.
The MR-LDSRG(2) equations are written as a set of iterative equations:
\begin{align}
    \tens{t}{a}{i, \rm new} (s) &= [\tens{\bar{H}}{a}{i,\rm old} (s) + \tens{t}{a}{i,\rm old} (s) \tens{\Delta}{a}{i} ] \frac{1-e^{-s(\tens{\Delta}{a}{i})^2}}{\tens{\Delta}{a}{i}}, \label{eq:t1_update}\\
    \tens{t}{ab}{ij, \rm new} (s) &= [\tens{\bar{H}}{ab}{ij,\rm old} (s) + \tens{t}{ab}{ij,\rm old} (s) \tens{\Delta}{ab}{ij} ] \frac{1-e^{-s(\tens{\Delta}{ab}{ij})^2}}{\tens{\Delta}{ab}{ij}}, \label{eq:t2_update}
\end{align}
which are solved using as a starting guess first-order amplitudes obtained from a DSRG-MRPT2 computation.\cite{Li:2015iz}

Matrix elements of the one- and two-body DSRG transformed Hamiltonians [Eqs.~\eqref{eq:t1_update} and \eqref{eq:t2_update}] are computed by accumulating the nested commutators in Eq.~\eqref{eq:BCH}:
\begin{equation}
\label{eq:recursive_sum}
\bar{H}_{1,2} = \sum_{k=0}^\infty \hat{O}_{k} (s),
\end{equation}
where the $k$th-nested term [$\hat{O}_{k} (s)$] is  obtained from the recursive equation:
\begin{align}
\label{eq:recursive}
    \hat{O}_{k} (s) &= \frac{1}{k} [\hat{O}_{k-1} (s), \hat{A}(s)]_{1,2}, \quad k = 1, 2, 3, \cdots,
\end{align}
starting from $\hat{O}_{0} = \hat{H}$.
In Appendix \ref{appendix:matrix_elements}, we report all equations to compute the commutator $\hat{O} (s)~=~[\hat{H}, \hat{A} (s)]_{1,2}$, which is sufficient to obtain $\bar{H}(s)_{1,2}$ via Eq.~\eqref{eq:recursive_sum}.

The MR-LDSRG(2) equations for the energy and amplitudes consist of 39 terms (in the spin orbital formalism).
In comparison, SR CC theory with singles and doubles requires 48 diagrams in total, while the ic-MRCC equations have a significantly larger number of terms.\cite{Hanauer2011p204111}
For small active spaces, the computational cost of MR-LDSRG(2) is dominated by the contribution:
\begin{equation}
\tens{O}{pb}{qj} (s) \leftarrow
 \sum_{a}^{\mathbf{P}}  \sum_{ik}^{\mathbf{H}} \tens{v}{kp}{aq} \tens{t}{ab}{ij} (s) \density{k}{i} , \quad j \in \mathbf{H}, b \in \mathbf{P}, p,q \in \mathbf{G},
\end{equation}
which, after factorization, has a computational cost that scales as ${\cal O}(N^2 N_{\rm P}^2 N_{\rm H}^2)$.
The term with the worst scaling with respect to the number of active orbitals has a cost of ${\cal O}(N_{\rm A}^6 N_{\rm V})$, which it is still significantly cheaper that the cost required by the orthonormalization step in projective theories [${\cal O}(N_{\rm A}^9)$].

In the MR-DSRG, reference relaxation effects are accounted for by solving the eigenvalue equation [Eq.~\eqref{eq:eigen}].
To diagonalize the $\bar{H}_{1,2}(s)$ within the space of reference determinants, we first express $\bar{H}(s)$ using second quantized operators that are normal ordered with respect to the true vacuum.
Specifically, we write the one- and two-body terms of $\bar{H} (s)$ as:
\begin{equation}
\begin{split}
    \bar{H}_{1,2}(s) =& \bar{H}_0 (s)   - \sum_{ij} \tens{\bar{H}}{i}{j} (s) \density{i}{j}\\
    &+ \frac{1}{2} \sum_{ijkl} \tens{\bar{H}}{ij}{kl} (s) \density{i}{k} \density{j}{l} - \frac{1}{4} \sum_{uvxy} \tens{\bar{H}}{uv}{xy} (s) \cumulant{uv}{xy} \\
    &+\sum_{pq} \Big[ \tens{\bar{H}}{p}{q} (s)  - \sum_{ij} \tens{\bar{H}}{pi}{qj} (s) \density{i}{j}  \Big] \sqop{p}{q}, \\
    &+ \frac{1}{4} \sum_{pqrs} \tens{\bar{H}}{pq}{rs} (s) \sqop{pq}{rs}.
\end{split}
\end{equation}
When the $\bar{H}_{1,2}(s)$ is written in this form, the quantities $[\tens{\bar{H}}{p}{q} (s) - \sum_{ij}\tens{\bar{H}}{pi}{qj} (s) \density{i}{j}]$ and $[\tens{\bar{H}}{pq}{rs} (s)]$ may be readily identified as MR-DSRG dressed one- and two-electron integrals, respectively.
These quantities can be used to build and diagonalize $\bar{H}(s)$ in the CASCI space and determine the density matrix and cumulants for the new reference.
Practically, we find that 5--10 macroiterations are required to converge the energy to less than $10^{-8}$ \Eh.

\section{Computational Details}
\label{sec:comput_detail}

The ground-state singlet potential energy curves (PECs) of HF and N$_2$ were computed using the MR-LDSRG(2), Mukherjee MRCC theory with singles and doubles (Mk-MRCCSD),\cite{Mukherjee1998p157,Mahapatra1998p163,Mukherjee1999p6171,Evangelista2006p154113,Evangelista2007p024102} MRCI with singles and doubles (MRCISD),\cite{Knowles1988p514,Knowles1988p5803} MRCISD with Davidson correction\cite{Davidson1974p61} (MRCISD+Q), and full configuration interaction (FCI).
Special treatments were applied to the Mk-MRCCSD computations of N$_2$:
(1) Tikhonov regularization\cite{Bartlett2009p144112} ($\omega = 0.01$) was used throughout the iterations to aid convergence;
(2) the effective Hamiltonian matrix elements between determinants that differ by more than two spin orbitals were neglected.
Spectroscopic constants of HF and N$_2$ were obtained by fitting the PECs with a ninth-order polynomial centered around the equilibrium geometry and compared to results from coupled cluster theory with singles and doubles (CCSD),\cite{Bartlett1982p1910} CCSD with perturbative triples [CCSD(T)],\cite{Pople1989p479,Stanton1997p130} and unitary DSRG with one- and two-body operators [DSRG(2)].\cite{Evangelista2014p054109}

The singlet-triplet splitting of \textit{para}-benzyne was studied using the MR-LDSRG(2) theory in combination with two active spaces: CAS($2, 2$) and CAS($8, 8$).
The former consists of two carbon $\sigma$ orbitals on radical centers, while the latter further includes six carbon $\pi$ orbitals.
Optimized geometries of singlet and triplet \textit{p}-benzynes computed at the Mk-MRCCSD/cc-pVTZ level of theory using a CASSCF($2,2$) reference were taken from Ref.~\citenum{Gauss2012p204108}.

All computations utilized Dunning's correlation consistent double-$\zeta$ (cc-pVDZ) basis set\cite{Dunning1989p1007} and semicanonical CASSCF orbitals, obtained by diagonalizing the core, active, and virtual blocks of the generalized Fock matrix.
Carbon, nitrogen, and fluorine 1s core orbitals were allowed to relax in the CASSCF computations, but were frozen in all subsequent treatments of electron correlation.
We used the \molpro 2015.1 package\cite{MOLPRO-WIREs,MOLPRO2015} to obtain the MRCISD and FCI energies, and the \PSI program\cite{PSI4} for the remaining computations.
All FCI energies are provided in the supplementary material.
\section{Results}
\label{sec:results}

\subsection{Hydrogen fluoride, CAS(2,2)}
\label{sec:HF}
 
\begin{figure}[!ht]
\centering
    \includegraphics[width=\columnwidth]{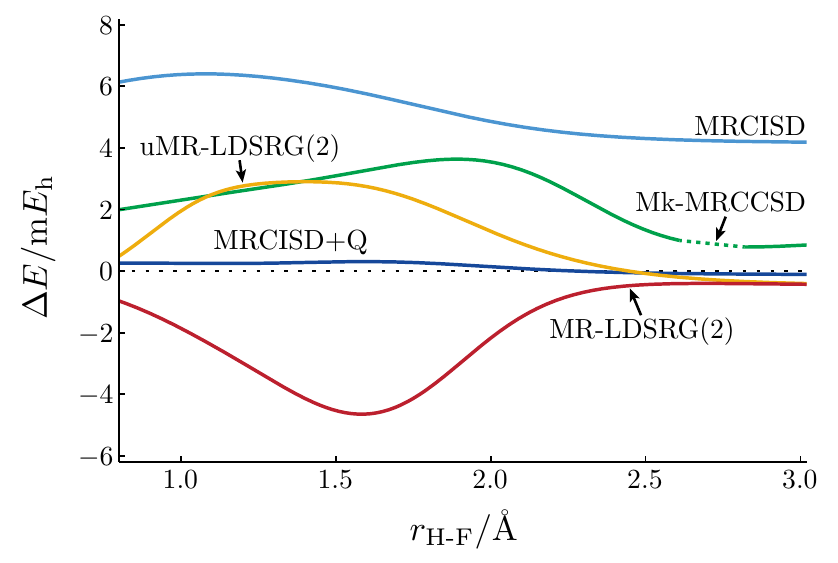}
    \caption{Energy deviations relative to FCI for the $X\,^1\Sigma^+$ state of HF computed using various multireference methods based on a CASSCF($2,2$) reference and the cc-pVDZ basis set. All MR-LDSRG(2) curves are computed using $s = 0.5$ \sunit. 
    The dashed line indicates the range of the plot for which Mk-MRCCSD computations failed to converge (2.60--2.85 \AA).}
    \label{fig:curve_HF}
\end{figure}

To investigate the ability of the MR-LDSRG(2) approach to describe single-bond breaking process, we study the ground-state dissociation curve of HF ($X\,^1 \Sigma^{+}$).
Figure~\ref{fig:curve_HF} presents the energy differences relative to the FCI of several multireference theories as a function of the bond distance ($r_\text{H-F}$).
For the MR-LDSRG(2) method, we report the energy computed with both an unrelaxed  and a fully relaxed reference (the former indicated with the prefix ``u'').
In all MR-LDSRG(2) calculations the flow variable is set equal to $s = 0.5$ \sunit, a value that has been shown to provide reliable results at the second-order perturbation level.\cite{Li:2015iz}

A comparison of the MR-LDSRG(2) curves shows that reference relaxation effects play a significant role at equilibrium and in the recoupling region ($r_\text{H-F} \in [1,2]$ \AA).
At long distances ($r_\text{H-F} > 2.5$ \AA), relaxation effects vanish because the reference coefficients are determined by symmetry, and as a result, both the relaxed and unrelaxed calculations converge to the same limit.
Judged from the nonparallelity error (NPE)---defined as the difference between the maximum and minimum signed errors---the unrelaxed (3.36 m\Eh) and relaxed (4.24 m\Eh) versions of the MR-LDSRG(2) yield curves that have slightly larger errors than those computed with  the MRCISD (2.24 m\Eh) and Mk-MRCCSD (2.85 m\Eh) methods.

In Table~\ref{tab:spec_cons_HF} we compare the equilibrium bond length ($r_{\rm e}$), harmonic vibrational frequency ($\omega_{\rm e}$), and the anharmonicity constant ($\omega_{\rm e}x_{\rm e}$) of HF ($X\,^1 \Sigma^+$) computed with various single-reference and multireference methods.
To gauge the $s$ dependence of the MR-LDSRG(2) results we consider both the case $s=0.5$ and 1.0 \sunit.
For the \uLDSRG, the change of $s$ causes a large shift in the value of equilibrium properties.  This is demonstrated, for example,  by the 22.3 \cm variation in the harmonic vibrational frequency.
As observed in the PECs calculations, properties computed with the relaxed MR-LDSRG(2) are less sensitive to the choice of $s$.
The shift in harmonic vibrational frequency is only 6.7 \cm, less than three times the value obtained with the unrelaxed approach.
In general, properties computed with the \uLDSRG and MR-LDSRG(2) methods are less accurate than those from SR-CC methods, Mk-MRCCSD, and MRCISD.

\newcolumntype{d}[1]{D{.}{.}{#1}}
\begin{table}[!ht]
\begin{threeparttable}
\centering
\renewcommand{\arraystretch}{1.25}
\caption{Spectroscopic constants for the $X\,^1 \Sigma^+$ state of HF computed using various single-reference and multireference methods.  All computations use the cc-pVDZ basis set.  Coupled cluster calculations use a restricted Hartree--Fock reference, whereas multireference calculations are based on a CASSCF(2,2) reference.  All values are deviations from FCI results.}
\label{tab:spec_cons_HF}
\begin{tabular*}{\columnwidth}{@{\extracolsep{\stretch{1}}}ld{2.4}d{3.1}d{2.1}@{}}
    \hline

    \hline
    Method & \multicolumn{1}{c}{$r_e$/\AA} & \multicolumn{1}{c}{$\omega_e$/\cm} & \multicolumn{1}{c}{$\omega_ex_e$/\cm} \\
    \hline
    CCSD & -0.0014 & 25.8 & -1.4 \\
    CCSD(T) & -0.0004 & 7.0 & -0.2 \\
    DSRG(2) ($s = 1.0$) & 0.0009 & -21.9 & 1.7 \\
    CASSCF($2,2$) & 0.0008 & -81.3 & 10.2 \\
    DSRG-MRPT2 ($s = 0.5$) & -0.0026 & 10.3 & 1.4 \\
    DSRG-MRPT2 ($s = 1.0$) & -0.0065 & 15.0 & 6.4 \\
    \uLDSRG ($s = 0.5$) & -0.0035 & 50.3 & 6.0 \\
    \uLDSRG ($s = 1.0$) & -0.0041 & 72.6 & 18.0 \\
    MR-LDSRG(2) ($s = 0.5$) & 0.0021 & -40.0 & 0.8 \\
    MR-LDSRG(2) ($s = 1.0$) & 0.0022 & -33.3 & 3.7 \\
    Mk-MRCCSD & -0.0008 & 11.0 & 0.0 \\
    MRCISD & -0.0005 & 1.0 & 0.8 \\
    MRCISD+Q & 0.0000 & -0.8 & -0.1 \\
    FCI & 0.9203 & 4143.2 & 92.9 \\
    \hline

    \hline
\end{tabular*}
\end{threeparttable}
\end{table}

\subsection{Nitrogen molecule, CAS(6,6)}
\label{sec:N2}

\begin{table*}[ht!]
\begin{threeparttable}
\centering
\renewcommand{\arraystretch}{1.25}
\caption{Energy errors (in m\Eh) for N$_2$ ($X\,^1 \Sigma_{g}^{+}$) at several atomic distances ($r_\text{N-N}$, in \AA). All computations used a CASSCF($6,6$) reference and the cc-pVDZ basis set.  Correlated methods included only \emph{single} and \emph{double} excitations and employed the frozen-core approximation.}
\label{tab:energetics_N2}
\footnotesize

\ifpreprint
\begin{tabular*}{\columnwidth}{@{\extracolsep{\stretch{1}}}*{1}{d{1.4}}*{2}{d{2.3}}*{2}{d{1.3}}*{3}{d{1.3}}*{2}{d{2.3}}*{1}{d{2.3}}*{1}{d{4.6}}@{}}
\else
\begin{tabular*}{2\columnwidth}{@{\extracolsep{\stretch{1}}}*{1}{d{1.4}}*{2}{d{2.3}}*{2}{d{1.3}}*{3}{d{1.3}}*{2}{d{2.3}}*{1}{d{2.3}}*{1}{d{4.6}}@{}}
\fi
\hline

\hline
 & & & & & \multicolumn{3}{c}{MR-LDSRG(2)} & & & & \\
 \cline{6-8}
 &  &  &  &  & \multicolumn{1}{c}{unrelaxed} & \multicolumn{2}{c}{relaxed} &  &  &  &  \\
 \cline{6-6} \cline{7-8}
  \multicolumn{1}{c}{$r_\text{N-N}$} & \multicolumn{1}{c}{LCT\tnote{a}} & \multicolumn{1}{c}{L3CT\tnote{a}} & \multicolumn{1}{c}{QCT\tnote{a}} & \multicolumn{1}{c}{Q3CT\tnote{a}} & \multicolumn{1}{c}{$s = 0.5$} & \multicolumn{1}{c}{$s = 0.5$} & \multicolumn{1}{c}{$s = 1.0$} & \multicolumn{1}{c}{MRCI} & \multicolumn{1}{c}{MRCI+Q} & \multicolumn{1}{c}{Mk-MRCC\tnote{b}} & \multicolumn{1}{c}{FCI\tnote{c}} \\
\hline
0.9525 & -0.421 & -1.781 & 4.620 & 3.387 & 3.702 & 2.613 & 2.455  & 8.391 & -0.564 & 4.750 & -109.167\,573 \\
1.0679 & -0.576 & -2.575 & 5.191 & 3.257 & 4.819 & 3.287 & 2.829 & 8.883 & -0.782 & 6.576 & -109.270\,384 \\
1.1208 & -0.281 & -2.582 & 5.426 & 3.043 & 5.487 & 3.703 & 3.207 & 9.123 & -0.845 & 7.193 & -109.278\,339 \\
1.1737 & -0.178 & -2.696 & 5.818 & 2.913 & 6.147 & 4.113 & 3.613 & 9.348 & -0.812 & 7.874 & -109.271\,915 \\
1.2700 & 0.342 & -2.852 & 7.044 & 3.387 & 7.188 & 4.781 & 4.344 & 9.734 & -1.029 & 9.010 & -109.238\,397 \\
1.4288 & 0.613 & -2.993 & 7.745 & 3.280 & 8.171 & 5.466 & 5.272 & 10.313 & -1.174 & 11.341 & -109.160\,305 \\
1.5875 & 0.474 & -2.873 & 7.672 & 3.485 & 8.444 & 5.527 & 5.675 & 10.634 & -1.363 & 13.373 & -109.086\,211 \\
1.7463 & 0.07 & -2.00 & 6.34 & 3.47 & 8.81 & 5.57 & 6.08 & 10.66 & -1.49 & 15.72 & -109.030\,31 \\
1.9050 & -1.24 & -1.89 & 4.07 & 3.09 & 9.12 & 5.85 & 6.62 & 10.14 & -1.75 & 17.56 & -108.994\,81 \\
\multicolumn{1}{c}{NPE\tnote{d}} & 1.85 & 1.213 & 3.68 & 0.571 & 5.41 & 3.23 & 4.17 & 2.72 & 1.19 & 12.81 & \\
\hline

\hline
\end{tabular*}
\begin{tablenotes}
\item [a] From Ref.~\citenum{Neuscamman:2009cy}. L3CT and Q3CT include the exact three-body reduced density matrix.
\item [b] The Mk-MRCC effective Hamiltonian elements between determinants that differ by more than two spin orbitals are neglected.
\item [c] FCI absolute energies (in \Eh) taken from Ref.~\citenum{Larsen:2000bp}.
\item [d] Non-parallel error (NPE) computed using these nine points.
\end{tablenotes}
\end{threeparttable}
\end{table*}

\begin{figure}[!ht]
\centering
    \includegraphics[width=\columnwidth]{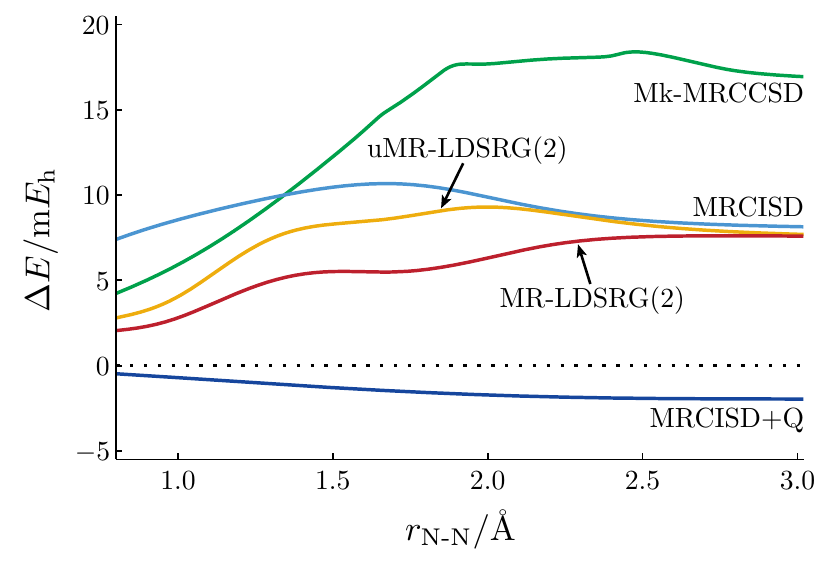}
    \caption{Energy deviations relative to FCI for the $X\,^1 \Sigma_{g}^{+}$ state of N$_2$ computed using various methods based on a CASSCF($6, 6$) reference and the cc-pVDZ basis set. All MR-LDSRG(2) curves used $s = 0.5$ \sunit. 
    The Mk-MRCCSD curve results from the approximated Mk-MRCCSD computations described in Sec.~\ref{sec:comput_detail}.}
    \label{fig:curve_N2}
\end{figure}

Next, we focus on the PEC for the $X\,^1 \Sigma_{g}^{+}$ state of N$_2$.
Energy errors with respect to FCI computed at various atomic distances ($r_\text{N-N}$) are summarized in Table~\ref{tab:energetics_N2} and plotted in Fig.~\ref{fig:curve_N2}.
\bibnote{Due to the high cost of FCI computations, the FCI PEC was generated on the grid of 0.1 \AA, while curves for other methods were constructed using a much finer grid (0.01 \AA).
To evaluate the error with respect to FCI on the finer grid, we perform a fourth-order least squares polynomial fit of the MRCISD+Q and FCI energy difference, $f(r) = E_\text{MRCISD+Q} (r) -  E_\text{FCI} (r) = \sum_{n=0}^{4} a_{n} r^{n}$.
Thus at a certain atomic distance $r$, the energy deviation for the method $X$ relative to FCI is calculated as $\Delta E(r) = E_X (r) - E_\text{MRCISD+Q} (r) + f(r)$.
}
In contrast to the case of hydrogen fluoride, for N$_2$ both the \uLDSRG and MR-LDSRG(2) methods are consistently in better agreement  with the reference with FCI curve than the MRCISD and Mk-MRCCSD approaches.
The NPEs of \uLDSRG and MR-LDSRG(2) are 5.25 and 4.81 m\Eh, respectively.
These results are comparable to the corresponding MRCISD number (3.27 m\Eh) and substantially smaller than the Mk-MRCCSD value (14.16 m\Eh).\bibnote{Notice that a similar NPE (12.03 m\Eh) is obtained for the Mk-MRCCSD implementation with full off-diagonal couplings using delocalized orbitals (see Ref.~\onlinecite{Kallay2010p074103})}

Table~\ref{tab:energetics_N2} also reports results for linear CT (LCT)\cite{Yanai2006p194106,Yanai2007p104107} and quadratic CT (QCT) theory, with and without the inclusion of the exact three-body density matrix.\cite{Neuscamman:2009cy}
The LCTSD scheme results are directly comparable to those from the MR-LDSRG(2) since both methods use the same commutator expansion and truncate the cluster operator to one- and two-body operators.
Interestingly, the LCTSD gives a NPE (1.85 m\Eh) that is smaller than the relaxed MR-LDSRG(2) value (3.23 m\Eh, $s = 0.5$ \sunit).
Another significant fact, is that the QCTSD approach---which uses an improved commutator expansion---gives a NPE (3.68 m\Eh) larger than the approaches based on a linearized commutator approximation.
This observation can be explained by an analysis of the errors introduced by truncating nested commutators up to two-body operators.\cite{Evangelista2012p27}
We also note that the inclusion of the three-body density matrix improves the performance of both LCT and QCT, but increases the computational scaling to $\mathcal{O}(N^7)$ and $\mathcal{O}(N^8)$, respectively.

Another interesting comparison can be made between MR-LDSRG(2) and the strongly contracted (SC) and weakly contracted (WC) versions of CT.\cite{Neuscamman:2010dz}
Both the SC- and WC-CTSD methods have a computational complexity analogous to that of the MR-LDSRG(2) approach, as they avoid diagonalizing the semi-internal excitation overlap metric [a ${\cal O}(N_{\rm A}^9)$ step].
The N$_2$ data summarized in Table~\ref{tab:dsrg_vs_ct} show that the MR-LDSRG(2) scheme yields results of quality intermediate between that of the WC- and SC-CTSD methods.
However, note that these two variants of CT are affected by the intruder-state problem and that some of the results reported in Table~\ref{tab:dsrg_vs_ct} were obtained by manually removing excitations linked to intruders.\cite{Neuscamman:2010dz}

\begin{table}[!ht]
\begin{threeparttable}
\centering
\renewcommand{\arraystretch}{1.25}
\caption{Errors (m\Eh) relative to MRCISD+Q for the $X\,^1 \Sigma_{g}^{+}$ state of N$_2$ in the range of $1.0 \leq r_\text{N-N} \leq 3.0$ \AA. All computations used a CASSCF($6,6$) reference and the cc-pVDZ basis set.  Core orbitals were not correlated.}
\label{tab:dsrg_vs_ct}
\begin{tabular*}{\columnwidth}{@{\extracolsep{\stretch{1}}}ld{2.3}d{1.3}d{1.3}d{1.3}d{2.3}@{}}
    \hline

    \hline
    & \multicolumn{3}{c}{MR-LDSRG(2)} & \multicolumn{2}{c}{LCTSD\tnote{a}} \\
    \cline{2-4} \cline{5-6}
    & \multicolumn{1}{c}{unrelaxed} & \multicolumn{2}{c}{relaxed} &  &  \\
    \cline{2-2} \cline{3-4}
    Error & \multicolumn{1}{c}{$s = 0.5$} & \multicolumn{1}{c}{$s = 0.5$} & \multicolumn{1}{c}{$s = 1.0$} & \multicolumn{1}{c}{SC} & \multicolumn{1}{c}{WC} \\
    \hline
    MIN & 4.753 & 3.506 & 3.191 & 5.280 & 3.524 \\
    MAX & 11.026 & 9.563 & 10.701 & 9.206 & 11.096 \\
    NPE & 6.273 & 6.057 & 7.510 & 3.927 & 7.977 \\
    \hline

    \hline
\end{tabular*}
\begin{tablenotes}
\item [a] From Ref.~\citenum{Neuscamman:2010dz}.
\end{tablenotes}
\end{threeparttable}
\end{table}

Table~\ref{tab:spec_cons_N2} reports the spectroscopic constants for the ground state of N$_2$.
Contrary to the case of HF, all MR-LDSRG(2) methods yield results comparable to those of the approximated Mk-MRCCSD and the single reference DSRG(2), and considerably exceed the quality of the CCSD results.
The MR-LDSRG(2) method provides the most reliable predictions, which differ from FCI by 0.0016 \AA{} ($r_{\rm e}$), 15.4 \cm ($\omega_{\rm e}$), and 0.2 \cm ($\omega_{\rm e}x_{\rm e}$).
Another encouraging observation is that both going from a perturbative to a nonperturbative treatment of dynamic correlation and the inclusion of relaxation effects contribute to reducing the $s$ dependence of the MR-DSRG methods.

\begin{table}[!ht]
\begin{threeparttable}
\centering
\renewcommand{\arraystretch}{1.25}
\caption{Spectroscopic constants for the $X\,^1 \Sigma_g^+$ state of N$_2$ computed using various single-reference and multireference methods.  All computations use the cc-pVDZ basis set.  Coupled cluster calculations use a restricted Hartree--Fock reference, whereas multireference calculations are based on a CASSCF(6,6) reference.  All values are deviations from FCI results.}
\label{tab:spec_cons_N2}
\begin{tabular*}{\columnwidth}{@{\extracolsep{\stretch{1}}}ld{2.4}d{3.1}d{2.1}@{}}
    \hline

    \hline
    Method & \multicolumn{1}{c}{$r_e$/\AA} & \multicolumn{1}{c}{$\omega_e$/\cm} & \multicolumn{1}{c}{$\omega_ex_e$/\cm} \\
    \hline
    CCSD & -0.0073 & 85.2 & -1.4 \\
    CCSD(T) & -0.0012 & 15.3 & -0.4 \\
    DSRG(2) ($s = 1.0$) & -0.0013 & 35.5 & -2.7 \\
    CASSCF($6,6$) & -0.0058 & 41.8 & -0.3 \\
    DSRG-MRPT2 ($s = 0.5$) & -0.0011 & -3.5 & 0.7 \\
    DSRG-MRPT2 ($s = 1.0$) & -0.0019 & 7.8 & 1.1 \\
    \uLDSRG ($s = 0.5$) & -0.0025 & 21.8 & 0.4 \\
    \uLDSRG ($s = 1.0$) & -0.0027 & 29.1 & 0.4 \\
    MR-LDSRG(2) ($s = 0.5$) & -0.0016 & 13.7 & 0.2 \\
    MR-LDSRG(2) ($s = 1.0$) & -0.0015 & 15.4 & -0.0 \\
    Mk-MRCCSD\tnote{a} & -0.0022 & 22.9 & -0.2 \\
    MRCISD & -0.0009 & 6.5 & -0.0 \\
    MRCISD+Q & 0.0002 & -2.2 & 0.0 \\
    FCI & 1.1201 & 2323.6 & 14.9 \\
    \hline

    \hline
\end{tabular*}
\begin{tablenotes}
\item [a] The Mk-MRCC effective Hamiltonian elements between determinants that differ by more than two spin orbitals are neglected.
\end{tablenotes}
\end{threeparttable}
\end{table}

\subsection{\textit{p}-Benzyne, CAS(2,2) and CAS(8,8)}
\label{sec:benzyne}

\begin{table}[!ht]
\begin{threeparttable}
\centering
\ifpreprint
\renewcommand{\arraystretch}{0.8}
\else
\renewcommand{\arraystretch}{1.25}
\fi
\caption{Adiabatic singlet-triplet splittings ($\Delta E_{\rm ST} = E_{\rm T} - E_{\rm S}$, in \kcal) of \textit{p}-benzyne computed using various multireference methods and the cc-pVDZ basis set. All computational results include a zero-point vibrational energy (ZPVE) correction equal to $+$0.30 \kcal.  Geometries and the ZPVE correction are taken from Ref.~\citenum{Gauss2012p204108}.}
\label{tab:benzyne_Est}
\begin{tabular*}{\columnwidth}{@{\extracolsep{\stretch{1}}}cld{1.2}@{}}
    \hline

    \hline
    Active Space & Method & \multicolumn{1}{c}{$\Delta E_\text{ST}$} \\
    \hline
    \multirow{12}{*}{CAS($2,2$)} & CASSCF & 0.27 \\
    & DSRG-MRPT2 ($s = 0.5$) & 2.55 \\
    & \uLDSRG ($s = 0.5$) & 2.15 \\
    & \uLDSRG ($s = 1.0$) & 2.72 \\
    & MR-LDSRG(2) ($s = 0.5$) & 3.51 \\
    & MR-LDSRG(2) ($s = 1.0$) & 5.32 \\
    & MRCISD & 1.75 \\
    & MRCISD+Q & 2.67 \\
    & Mk-MRCCSD & 5.23 \\
    & Mk-MRCCSD(T) & 4.49 \\
    & ic-MRCCSD\tnote{a} & 4.02 \\
    & ic-MRCCSD(T)\tnote{a} & 5.06 \\
    \hline
    \multirow{11}{*}{CAS($8,8$)} & CASSCF & 2.37 \\
    & DSRG-MRPT2 ($s = 0.5$) & 4.22 \\
    & \uLDSRG ($s = 0.5$) & 4.04 \\
    & \uLDSRG ($s = 1.0$) & 4.46 \\
    & MR-LDSRG(2) ($s = 0.5$) & 4.71 \\
    & MR-LDSRG(2) ($s = 1.0$) & 5.50 \\
    & MRCISD & 3.54 \\
    & MRCISD+Q & 4.18 \\
    & Mk-MRCCSD\tnote{b} & 5.23 \\
    & Mk-MRCCSD(T)\tnote{b} & 3.86 \\
    & ic-MRCCSD\tnote{a} & 4.95 \\
    & ic-MRCCSD(T)\tnote{a} & 5.25 \\
    \hline
    Experiment\tnote{c} & & \multicolumn{1}{c}{$3.8 \pm 0.4$} \\
    \hline

    \hline
\end{tabular*}
\begin{tablenotes}
\item[a] From Ref.~\citenum{Andreas2012p204107}.
\item[b] Approximated value obtained using the same procedure for N$_2$ described in Sec.~\ref{sec:comput_detail}.
\item[c] Ultraviolet photoelectron spectroscopy from Ref.~\citenum{Lineberger1998p5279}.
\end{tablenotes}
\end{threeparttable}
\end{table}

In our final test case we use the MR-LDSRG(2) to compute the adiabatic singlet-triplet splitting ($\Delta E_{\rm ST} = E_{\rm T} - E_{\rm S}$) of \textit{p}-benzyne.\cite{Wenk:2003wv,Cramer1997p311,Lindh1999p9913,Crawford:2001we,Krylov2002p4694,Li2007p12,Evangelista2007p024102,Parish2008p044306,Paldus2008p174101,Gauss2012p204108,Andreas2012p204107,Schutski:2014it}
Our reference value was taken from the photoelectron spectroscopy experiments of Wenthold, Squires, and Lineberger.\cite{Lineberger1998p5279}
These authors obtained the value $\Delta E_{\rm ST}$ = 3.8 $\pm$ 0.5 \kcal, but also considered an alternative (but less likely) value of 2.1 \kcal.
%

Table~\ref{tab:benzyne_Est} reports the DSRG-MRPT2 and MR-LDSRG(2) singlet-triplet splitting computed with the cc-pVDZ basis set.  All results are shifted by $+0.30$ \kcal to account for zero-point vibrational energy (ZPVE) corrections.\cite{Gauss2012p204108}
The singlet-triplet splitting computed with the \uLDSRG method shows a marked dependence on the size of the active space and the error is dominated by the CASSCF contribution.
This can be seen from the fact that the correlation energy contribution to the splitting (e.g. $\Delta E_{\rm ST}^\text{MR-LDSRG(2)} - \Delta E_{\rm ST}^\text{CASSCF}$)  is almost the same for the CAS(2,2) and CAS(8,8) references.  For example, at $s=0.5$ \sunit, the correlation energy contribution to the splitting is 1.88 and 1.67 \kcal, respectively.
After introducing reference relaxation, the active space dependence is greatly alleviated, and becomes smaller as the flow parameter increases. 
For the MR-LDSRG(2) at $s = 1$ \sunit, the difference between $\Delta E_{\rm ST}$ computed with the CAS(2,2) and CAS(8,8) references is only 0.18 \kcal.

Our best estimates of $\Delta E_{\rm ST}$ computed using the MR-LDSRG(2) based on a CASSCF(8,8) reference are 4.71 and 5.50 \kcal for $s=0.5$ and 1.0  \sunit, respectively.
These values are in good agreement with the ic-MRCCSD and ic-MRCCSD(T) results computed with the largest active space: 4.95 and 5.25 \kcal, respectively.
Notice that $\Delta E_{\rm ST}$ from MRCISD and MRCISD+Q shows a marked dependence on the size of the active space, while the ic-MRCCSD and ic-MRCCSD(T) results display smaller variations.
Interestingly, the CAS(8,8) Mk-MRCCSD(T) singlet-triplet splitting (3.86 m\Eh) is the one that comes the closest to the experimental value (3.8 m\Eh).
This result is likely to be fortuitous, since the quality of the Mk-MRCC approach is known to degrade as the active space is increased.\cite{Kong:2010ek,Gauss2013p176}
Another issue to take into consideration is the fact that Mk-MRCC computations have a cost proportional to the number of reference determinants, which makes this approach impractical for large active spaces.
Indeed, our \textit{p}-benzyne CAS(8,8) Mk-MRCCSD computations cost about 660 times more than a single CCSD calculation.

\subsection{Evolution of the MR-LDSRG(2) flow}
\label{sec:evo}

\begin{figure}
\centering
    \includegraphics[width=\columnwidth]{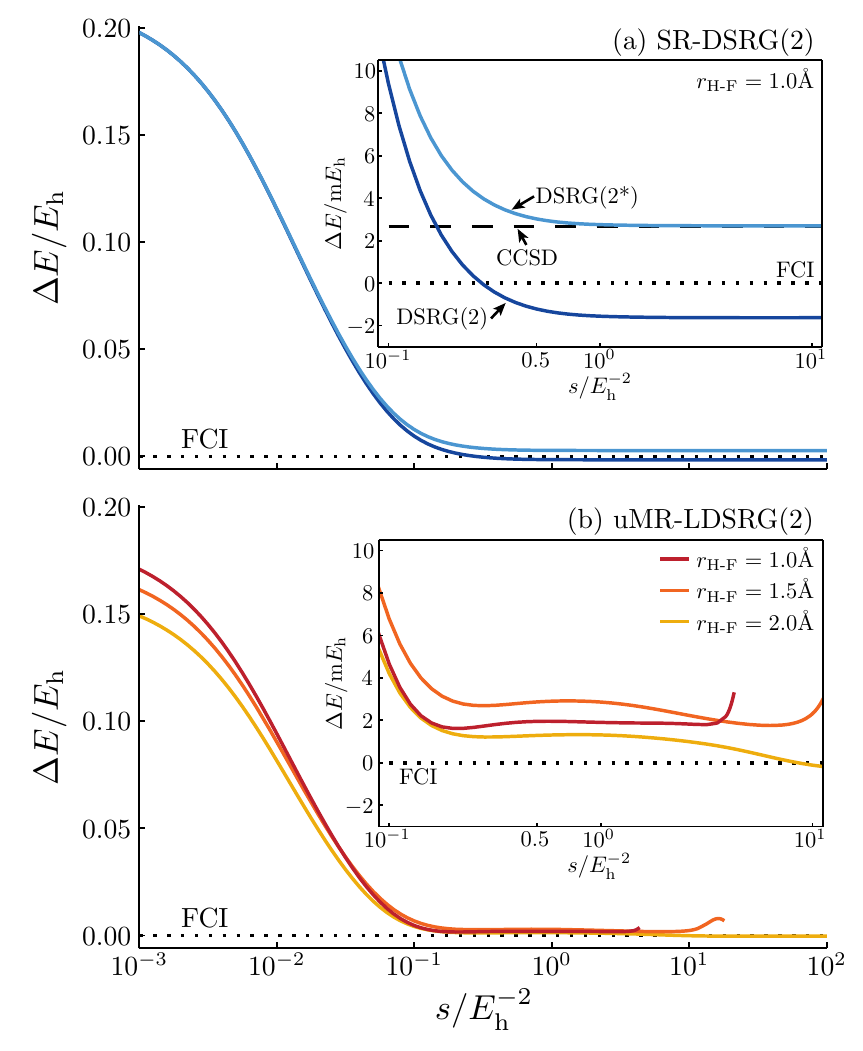}
    \caption{Energy deviations relative to FCI for HF ($X\,^1 \Sigma^{+}$) plotted against the flow parameter $s$ (in logarithm scale): (a) SR-DSRG(2) at $r_\text{H-F} = 1.0$ \AA, and (b) \uLDSRG at $r_\text{H-F} = 1.0, 1.5$, and $2.0$ \AA. The insets show the corresponding enlarged plots for $0.1 \leq s \leq 10.0$ \sunit. All computations used the cc-pVDZ basis set.}
    \label{fig:s_evo}
\end{figure}

In this section we analyze the evolution of the MR-LDSRG(2) energy as a function of the flow variable $s$.  To this end, we consider the ground state of hydrogen fluoride at three bond lengths: 1.0, 1.5, and 2.0 \AA.
Figure~\ref{fig:s_evo} depicts the energy errors of both SR- and MR-DSRG with respect to the FCI as a function of $s$.
Specifically, we consider the unrelaxed \uLDSRG, the single-reference (SR) DSRG(2), and the fourth-order energy corrected version of the DSRG(2) [DSRG(2*)].\cite{Evangelista2014p054109}

The top panel of Figure~\ref{fig:s_evo} shows that energy error of the single-reference DSRG(2) and DSRG(2*) methods are monotonically decreasing functions of $s$.
This behavior is consistent with the flow of the energy in the similarity renormalization group (SRG).\cite{Hergert:2013kf}
In the limit of $s$ that goes to infinity, the DSRG(2*) energy is almost indistinguishable from the CCSD value, while the DSRG(2) overestimates the correlation energy.

On the contrary, the MR-LDSRG(2) energy does not decrease monotonically with respect to $s$.
This behavior was already observed in results from second-order MR-DSRG perturbation theory\cite{Li:2015iz} and applications of the \textit{in medium} multireference SRG to nuclear structure problems.\cite{Hergert:2013hc}
For large values of $s$, the MR-LDSRG(2) fails to converge when $r_\text{H-F} = 1.0$, and 1.5 \AA, while there are no issues at 2.0 \AA{}.
Convergence problems for large values of $s$ are expected, and can be understood by means of a perturbative analysis of the MR-LDSRG(2) equations.
The first-order MR-DSRG amplitudes for doubles are given by:\cite{Li:2015iz}
\begin{equation}
\label{eq:first_order_t2}
\tens{t}{ab}{ij,(1)}(s) = \frac{\tens{v}{ab}{ij} [1 - e^{-s(\epsilon_i + \epsilon_j - \epsilon_a - \epsilon_b)^2}]}
{\epsilon_i + \epsilon_j - \epsilon_a - \epsilon_b}.
\end{equation}
In the limit of $s\rightarrow\infty$, the first-order MR-DSRG amplitudes are equivalent to the first-order M{\o}ller--Plesset amplitudes, and they diverge when the energy denominator $\epsilon_i + \epsilon_j - \epsilon_a - \epsilon_b$ approaches zero.
The inset of Fig.~\ref{fig:s_evo}(b) shows details of the MR-LDSRG(2) energy for $s$ in the range $[0.1, 10]$ \sunit.
We notice that our recommended range for $s$, $[0.5, 1.0]$ \sunit, is located within an energy plateau, which is consistent with the observed weak $s$-dependence of our results.

Our experience with the single-reference DSRG\cite{Evangelista2014p054109} suggests that numerical instabilities may also be aggravated by the use of an approximate BCH expansion.
Indeed, when the linearized BCH approximation is modified to recover the correct prefactor for the leading third-order terms, the convergence of the resulting DSRG(2*) method is superior to that of the DSRG(2).
In fact, when we look at a different bond length ($r_\text{H-F} = 2.0$ \AA{}) than the one used in Fig.~\ref{fig:s_evo}(a), the DSRG(2) becomes numerically unstable for $s > 2.0$ \sunit, while the DSRG(2*) always converges in the sampled region ($s \leq 10^2$ \sunit).

\section{Formal comparison of the MR-DSRG with other multireference methods}
\label{sec:comparison}

In this section we will summarize the similarities and differences between the MR-DSRG formalism and other nonperturbative multireference theories.
Readers may immediately recognize the close connection between the MR-DSRG and canonical transformation (CT) theory of Yanai and Chan.\cite{Yanai2006p194106,Yanai2007p104107}
Both methods transform the Hamiltonian unitarily, and evaluate the BCH expansion using a recursive commutator approximation.\cite{Yanai2006p194106,Evangelista:2012hz}
However, there are several important distinctions between the MR-DSRG and LCTSD approaches.
Firstly, reference relaxation effects were not considered in the formulation of CT theory.
However, semi-internal excitations ($\sqop{mx}{uv}$ and $\sqop{xy}{ev}$) still allow some degree of indirect reference relaxation in CT theory.

Secondly, the MR-DSRG relies on a set of many-body equations, while the CT scheme uses a projective formalism.
More precisely, the CT amplitudes are determined from a set of generalized Brillouin conditions of the form:\cite{Kutzelnigg:1979ex,Mukherjee:2001ci}
\begin{align}
\label{eq:CT_BC}
    \braket{\mref| [e^{-\hat{A}}\hat{H}e^{\hat{A}}, \sqop{ij\cdots}{ab\cdots} - \sqop{ab\cdots}{ij\cdots}] |\mref} = 0,
\end{align}
where $\hat{A}$ is analogous to the MR-DSRG $\hat{A}(s)$ operator but does not depend on $s$ and it is normal-ordered with respect to the true vacuum.
Moreover, since the basis of states $\sqop{ab\cdots}{ij\cdots}\ket{\mref}$ is nonorthogonal and linearly dependent, in CT it is necessary to orthogonalize this basis.
The most demanding step of the orthogonalization procedure involves semi-internal excitations and scales as ${\cal O}(N_{\rm A}^9)$.
The MR-LDSRG(2) approach avoids orthogonalization of the excitation manifold by employing many-body conditions [Eq.~\eqref{eq:many-body}],\cite{Lindgren1978p33,Bartlett1996p2652,Nooijen2011p214116}
and as a result, it has a lower scaling with respect to the size of the active space.



Other approaches closely related to the MR-DSRG include the internally-contracted MRCC theory,\cite{Evangelista2011p114102,Hanauer2011p204111} the state-specific partially internally contracted MRCC (\textit{p}IC-MRCC)\cite{Nooijen2011p214116} and the MR equation-of-motion CC (MR-EOMCC) theory of Datta and Nooijen.\cite{Datta2012p204107,Demel:2013kz,Nooijen2014p081102,Huntington:2016fo}
As in the case of CT theory, the ic-MRCC formalism is projective, but it relies on a nonunitary transformation of the bare Hamiltonian and does allow for relaxation of the reference wave function.

The \textit{p}IC-MRCC and MR-EOMCC are two \textit{transform and diagonalize} approaches.
For example, in the MR-EOMCC method, the Hamiltonian is similarity transformed according to:
\begin{equation}
\hat{G} = \{ e^{\hat{Y}} \}^{-1} e^{-\hat{T}'} \hat{H} e^{\hat{T}'} \{e^{\hat{Y}}\},
\end{equation}
where $\hat{T}'$ contains excitations from $\mathbf{H}$ to $\mathbf{V}$, while $\hat{Y}$ contains the non-commuting components of the ic-MRCC excitation operator.
The use of normal ordered exponential operators\cite{Lindgren:1978io} [$\{\exp(\hat{Y})\}$] instead of the traditional exponential operator simplifies the algebraic structure of the MR-EOMCC equations.\cite{Mukherjee1997p561,Mukherjee1997p432,Shamasundar2009p174109,Mukherjee2010p234107,Mukherjee2013p62,Kutzelnigg:2010iu}
Both the \textit{p}IC-MRCC and MR-EOMCC use a hybrid set of residual conditions.
Single excitations $\hat{T}$ are obtained from a set of projected equations of the form $\braket{\mref| \sqop{i}{e} \hat{G} |\mref} = 0$, while doubles amplitudes are derived from a set of many-body conditions.\cite{Nooijen2011p214116}
This mixed scheme has the advantage that one needs to orthogonalize only the space of single excited configurations.
Once $\hat{G}$ is determined, it is subsequently diagonalized in a space of determinants that spans a small multireference configuration interaction wave function.
Thus, both the \textit{p}IC-MRCC and MR-EOMCC theories properly account for reference relaxation effects.

For reasons that vary from method to method, all approaches considered here require the elimination of a portion of the cluster amplitudes.
In CT and ic-MRCC theory, the orthonormalization of the basis of excitation operators uses a numerical threshold to identify amplitudes that are redundant.
In the case of \textit{p}IC-MRCC and MR-EOMCC, despite the use of many-body conditions for doubles, it is still necessary to discard some doubles amplitudes that correspond to weakly occupied active orbitals.\cite{Nooijen2011p214116,Datta2012p204107}
In contrast, the combination of many-body equations and renormalization of intruders allows the MR-DSRG to retain all amplitudes and, in principle, avoid discontinuities caused by the elimination of  excitations.


\section{Conclusions}
\label{sec:conclusions}

The framework of similarity renormalization group provides a general approach to create many-body theories that do not suffer from problems with small energy denominators.
In this work we take advantage of this strategy to formulate the MR-LDSRG(2) approach, a novel multireference theory that combines numerical robustness with an internally-contracted treatment of dynamical electron correlation effects that is comparable to that of the single-reference CCSD approach.

The MR-DSRG formalism addresses two major difficulties encountered in other nonperturbative multireference theories: 1) convergence issues linked to the intruder-state problem and 2) energy discontinuities that arise from the need to eliminate redundant wave function parameters.
The MR-DSRG performs a continuous unitary transformation of the Hamiltonian that folds in dynamical correlation effects.
This transformation produces a flow renormalization of the many-body interaction, where problematic rotations between the reference and near-degenerate excited configurations are suppressed.\cite{Evangelista2014p054109}
The redundancy problem is dealt with a many-body formulation of the MR-DSRG equations,\cite{Lindgren1978p33,Bartlett1996p2652,Nooijen2011p214116} an approach that has been successfully applied to numerous MR methods.\cite{Nooijen2011p214116,Datta2012p204107,Demel:2013kz,Hergert:2013hc}
In addition, the MR-DSRG equations make extensive use of Mukherjee and Kutzelnigg's normal order formalism for multiconfigurational vacua.\cite{Mukherjee1997p561,Mukherjee1997p432,Shamasundar2009p174109,Mukherjee2010p234107,Mukherjee2013p62,Kutzelnigg:2010iu}

The MR-LDSRG(2) model introduced in this work is based on a cluster operator truncated to one- and two-body terms, while the Baker--Campbell--Hausdorff expansion is approximated with a linearized recursive formula. 
This model is perhaps one of the simplest internally contracted MR methods available: it contains only 39 terms and has a computational cost that scales as ${\cal O}(N^2 N_{\rm P}^2 N_{\rm H}^2)$, which is roughly the same as single reference CCSD [${\cal O}(N_{\rm P}^4 N_{\rm H})$].

The MR-LDSRG(s) has been benchmarked against the FCI ground-state potential energy curves (PECs) of HF and N$_2$, and the experimental singlet-triplet splitting of \textit{p}-benzyne.
The relaxed MR-LDSRG(2) PECs of HF and N$_2$ show similar nonparallelity errors, 4.24 m\Eh and 4.81 m\Eh, respectively, and maximum errors of comparable magnitude, 4.65 and 7.60 m\Eh, respectively.
To put these numbers into perspective, we also evaluate the CCSD and CCSD(T) dissociation energy of HF and N$_2$ as $D_e$(HF) = $E$(H,$^2S$) + $E$(F,$^2P$)$ - E$(HF,$r_e$) and $D_e$(N$_2$) = 2 $E$(N,$^4S$)$ - E$(N$_2$,$r_e$), respectively.
At the CCSD level, $D_e$(HF) and $D_e$(N$_2$) deviate from FCI by $-1.3$ and $-12.7$ m\Eh, respectively.
The addition of pertubative triples reduces these errors to $-0.3$ (HF) and $-1.6$ m\Eh (N$_2$).
Hence, the accuracy of the MR-LDSRG(2) appears to fall within the range expected for CCSD.
For \textit{p}-benzyne, the singlet-triplet gap is predicted to be 4.71 \kcal at the MR-LDSRG(2) ($s=0.5$) level of theory, a value that is within 1.2 \kcal from the experimentally measured gap and previously reported ic-MRCCSD and ic-MRCCSD(T) results.\cite{Andreas2012p204107}

We also notice that dependency of the MR-LDSRG(2) energy and properties on the value of the flow variable ($s$) is greatly reduced with respect to the DSRG second-order multireference perturbation theory (DSRG-MRPT2).\cite{Li:2015iz}
For example, when the flow variable $s$ is increased from 0.5 to 1.0 \sunit, the MR-LDSRG(2) equilibrium distances of HF and N$_2$ change by less than 0.0002 \AA, while at the DSRG-MRPT2 level they vary by 0.004 and 0.001 \AA, respectively.
Moreover, it is important to allow the reference wave function to relax in the presence of dynamic correlation, as shown by the conspicuous 1--3 \kcal changes in the \textit{p}-benzyne singlet-triplet splittings.
In general, we find that the relaxed MR-LDSRG(2) approach with $s = 0.5$ \sunit provides a consistent compromise between numerical robustness and accuracy.

Overall, our results suggest that future study should the natural next step 
would be to explore more accurate MR-DSRG truncation schemes.
Perhaps, the largest source of error in the MR-LDSRG(2) is the linear commutator approximation, since it is known to yield correlation energies that are correct only up to third order in perturbation theory.
One way to address this issue is to consider a quadratic commutator approximation.\cite{Neuscamman:2009cy}
Another aspect to consider is the inclusion of triple excitations via a perturbative correction analogous to the CCSD(T) approach.\cite{Pople1989p479}
In this respect, one of the advantages offered by the MR-DSRG formalism is that it does not require the costly orthogonalization of triple excitations, which is instead mandatory in methods that project equations onto a set of internally contracted configurations.

\begin{acknowledgments}
This work was supported by start-up funds provided by Emory University.
\end{acknowledgments}

\appendix
\section{MR-DSRG theory in a general basis}
\label{appendix:orbital_rotation}

As commented in Ref.~\citenum{Evangelista2014p054109}, the original formulation of the DSRG gives an energy that is not invariant with respect to separate rotations among orbitals that leave the reference unchanged (in the case of the MR-DSRG these are the core, active, and virtual orbitals), unless $s = 0$ or $s \rightarrow \infty$.
The lack of orbital invariance is due to the structure of the source operator [Eq.~\eqref{eq:source_tensor}].
The original parameterization of the source operator uses a Gaussian function of M{\o}ller--Plesset denominators in the semicanonical basis.
When orbitals are rotated to a different basis, the functional form of the source operator changes, thus, breaking orbital invariance.
By analyzing the issue of orbital invariance in the second-order SRG approach, we found a simple approach to write a general orbital-invariant DSRG source operator.
Without going in details of this derivation, our solution to the orbital invariance issue is to relate the source operator in an arbitrary basis to the original expression in the semicanonical basis via a series of unitary transformations.

To begin with, we need to establish the relationship between a set of general and semicanonical orbitals.
If we start from a noncanonical basis $\{\phi^{p}\}$, the unitary transformation $\psi^{p'} = \sum_p \tens{U}{p}{p'} \phi^p$ that connects it to the semicanonical basis $\{\psi^{p'}\}$ satisfies the eigenvalue problem for each block of the Fock matrix:
\begin{align}
    {\bf F}_X {\bf U}_{X} = {\bf U}_{X} \boldsymbol{\epsilon}_X, \quad X = {\rm C, A, V},
\end{align}
where ${\bf F}_X$ is the Fock matrix for block $X$ and $\boldsymbol{\epsilon}_X$ is the corresponding diagonal matrix of orbital energies.
The direct sum of these block transformations (${\bf U}_X$) yields the unitary matrix $({\bf U})$ that rotates a general basis to the semicanonical basis,
\begin{align}
    {\bf U} = {\bf U}_{\rm C} \oplus {\bf U}_{\rm A} \oplus {\bf U}_{\rm V}.
\end{align}
Following the notation of Kong,\cite{Kong:2010ek} we express the matrix element of $\bf U$ and its transpose as $U_{q}^{p'}$ and $U^{q}_{p'}$, respectively.

In a general basis obtained by rotating the semicanonical orbitals, the one- and two-body components of the source operator [Eq.~\eqref{eq:source_tensor}] can be rewritten as:
\begin{widetext}
\begin{align}
    \tens{r}{c}{k} &= \sum_{i'a'} U^{a'}_{c} [ ( \sum_{kc} U^{c}_{a'} \tens{\bar{H}}{c}{k} U_{k}^{i'} + \tens{\Delta}{a'}{i'} \sum_{kc} U^{c}_{a'} \tens{t}{c}{k} U_{k}^{i'} ) e^{-s (\tens{\Delta}{a'}{i'})^2} ] U^{k}_{i'}, \\
    \tens{r}{cd}{kl} &= \sum_{i'j'a'b'} U^{a'}_{c} U^{b'}_{d} [ ( \sum_{klcd} U^{d}_{b'} U^{c}_{a'} \tens{\bar{H}}{cd}{kl} U_{k}^{i'} U_{l}^{j'} + \tens{\Delta}{a'b'}{i'j'} \sum_{klcd} U^{d}_{b'} U^{c}_{a'} \tens{t}{cd}{kl} U_{k}^{i'} U_{l}^{j'} ) e^{-s (\tens{\Delta}{a'b'}{i'j'})^2} ] U^{l}_{j'} U^{k}_{i'}.
\end{align}
\end{widetext}
where the M{\o}ller--Plesset denominators $\tens{\Delta}{a'}{i'} = \dfock{i'} - \dfock{a'}$ and $\tens{\Delta}{a'b'}{i'j'} = \dfock{i'} + \dfock{j'} - \dfock{a'} - \dfock{b'}$, are defined in the semicanonical basis.

In practice, to evaluate the MR-DSRG equations, we first evaluate $\tens{\bar{H}}{}{\rm old}$ and $\tens{t}{}{\rm old}$ in a general basis, transformed them in the semicanonical basis, update the amplitudes using Eqs.~\eqref{eq:t1_update} and \eqref{eq:t2_update}, and transform the amplitudes back to the general basis.
The resulting algorithm is more expensive than directly solving the DSRG equation [Eq.~\eqref{eq:dsrg}] in the semicanonical basis since it requires additional steps that scale as $\mathcal{O}(N_{\rm H}^2N_{\rm P}^3)$.
Nevertheless, an orbital invariant formulation of the MR-DSRG allows us to evaluate the renormalized Hamiltonian in other bases that might offer a computational advantage (for example, the natural orbital basis).
We have implemented and numerically verified the orbital invariance of this new source operator on the singlet ground state of N$_2$.

\section{Matrix elements of $\bm{\hat{O} (s) = [\hat{H}, \hat{A} (s)]_{1,2}}$}
\label{appendix:matrix_elements}

Here we present the matrix elements of the linear commutator $\hat{O} (s) = [\hat{H}, \hat{A} (s)]_{1,2}$ required to evaluate the MR-DSRG transformed Hamiltonian via Eqs.~\eqref{eq:BCH} and \eqref{eq:recursive}.
Since $[\hat{H}, \hat{T}^\dag(s)] = - [\hat{H}, \hat{T}(s)]^\dag$ holds, only terms from $[\hat{H}, \hat{T}(s)]_{1,2}$ need to be derived.
As indicated by Eq.~\eqref{eq:bareH}, we may write the quantity $[\hat{H}, \hat{T}(s)]_{1,2}$ as the sum of four contributions,
\begin{align}
    [\hat{H}, \hat{T}(s)]_{1,2} =&[\hat{F}, \hat{T}_1(s)]_{1,2} + [\hat{F}, \hat{T}_2(s)]_{1,2} \notag\\
    &+ [\hat{V}, \hat{T}_1(s)]_{1,2} + [\hat{V}, \hat{T}_2(s)]_{1,2}.
\end{align}

\setcounter{table}{0}
\renewcommand{\thetable}{A\arabic{table}}
\begin{table*}[ht!]
\centering
\renewcommand{\arraystretch}{1.5}
\caption{Equations for the evaluation of the commutator $[\hat{H}, \hat{T} (s)]_{1,2}$ expressed using Einstein's notation.
Indices follow the convention introduced in Table~\ref{tab:orbital_space}.}
\label{tab:expression}

\ifpreprint
\begin{tabular*}{\columnwidth}{@{\extracolsep{\stretch{1}}}rll@{}}  
\else
\begin{tabular*}{2\columnwidth}{@{\extracolsep{\stretch{1}}}rll@{}}  
\fi

\hline
\hline
\# & Contribution & Expression \\
\hline

\multirow{2}{*}{1} & \multirow{2}{*}{$C$} & $+ \tens{f}{j}{b} \tens{t}{a}{i} \density{j}{i}\cdensity{a}{b} + \frac{1}{2} \cumulant{xy}{uv} [ ( \tens{f}{x}{e} \tens{t}{ey}{uv} + \tens{v}{xy}{ev} \tens{t}{e}{u} ) - ( \tens{f}{m}{v} \tens{t}{xy}{um} + \tens{v}{my}{uv} \tens{t}{x}{m} ) ] + \frac{1}{4} \tens{v}{kl}{cd} \tens{t}{ab}{ij} \density{k}{i} \density{l}{j} \cdensity{a}{c} \cdensity{b}{d}$ \\
& & $+ \cumulant{xy}{uv} \tens{v}{jx}{vb} \tens{t}{ay}{iu} \density{j}{i} \cdensity{a}{b} + \frac{1}{8} \cumulant{xy}{uv} ( \tens{v}{xy}{cd} \tens{t}{ab}{uv} \cdensity{a}{c} \cdensity{b}{d} + \tens{v}{kl}{uv} \tens{t}{xy}{ij} \density{k}{i} \density{l}{j} ) + \frac{1}{4} \cumulant{xyz}{uvw} ( \tens{v}{mz}{uv} \tens{t}{xy}{mw} + \tens{v}{xy}{we} \tens{t}{ez}{uv} )$ \\

\hline

2 & $\tens{C}{p}{i}$ & $+ \tens{f}{p}{a} \tens{t}{a}{i} + \frac{1}{2} \tens{v}{pw}{vy} \tens{t}{ux}{iz} \density{u}{v} \density{x}{y} \cdensity{w}{z} + \frac{1}{2} \tens{v}{pk}{cd} \tens{t}{ab}{ij} \density{k}{j} \cdensity{a}{c} \cdensity{b}{d} + \cumulant{xy}{uv} ( \frac{1}{4} \tens{v}{pj}{uv} \tens{t}{xy}{ij} + \tens{v}{px}{au} \tens{t}{ay}{iv} ) $ \\

3 & $\tens{C}{a}{p}$ & $- \tens{f}{i}{p} \tens{t}{a}{i} - \frac{1}{2} \tens{v}{ux}{pz} \tens{t}{aw}{vy} \cdensity{u}{v} \cdensity{x}{y} \density{w}{z} - \frac{1}{2} \tens{v}{kl}{pc} \tens{t}{ab}{ij} \density{k}{i} \density{l}{j} \cdensity{b}{c} - \cumulant{xy}{uv} ( \frac{1}{4} \tens{v}{xy}{pb} \tens{t}{ab}{uv} + \tens{v}{ix}{pu} \tens{t}{ay}{iv} )$ \\

4 & $\tens{C}{a}{i}$ & $+ \tens{t}{ab}{ij} \tens{f}{k}{b} \density{k}{j} - \tens{t}{au}{ij} \tens{f}{j}{v} \density{u}{v} + \frac{1}{2} \cumulant{xy}{uv} ( \tens{v}{yj}{uv} \tens{t}{ax}{ij} - \tens{v}{xy}{vb} \tens{t}{ab}{iu} )$ \\

5 & $\tens{C}{p}{q}$ & $+ \tens{t}{a}{i} \tens{v}{pj}{qa} \density{j}{i} - \tens{t}{u}{m} \tens{v}{pm}{qv} \density{u}{v} + \frac{1}{2} \cumulant{xy}{uv} ( \tens{v}{px}{qe} \tens{t}{ey}{uv} - \tens{v}{pm}{qu} \tens{t}{xy}{mv} )$ \\

\hline

6 & ${\cal P}(p,b)\, \tens{C}{pb}{ij}$ & $+ \tens{t}{ab}{ij} \tens{f}{p}{a}$ \\

7 & ${\cal P}(q,j)\, \tens{C}{ab}{jq}$ & $+ \tens{t}{ab}{ij} \tens{f}{i}{q}$ \\

8 & ${\cal P}(p,a)\, \tens{C}{ap}{rs}$ & $+ \tens{t}{a}{i} \tens{v}{pi}{rs}$ \\

9 & ${\cal P}(r,i)\, \tens{C}{pq}{ir}$ & $+ \tens{t}{a}{i} \tens{v}{pq}{ar}$ \\

10 & $\tens{C}{rs}{ij}$ & $+ \frac{1}{2} ( \tens{v}{rs}{cd} \tens{t}{ab}{ij} \cdensity{a}{c} \cdensity{b}{d} - \tens{v}{rs}{vy} \tens{t}{ux}{ij} \density{u}{v} \density{x}{y} )$ \\

11 & $\tens{C}{ab}{pq}$ & $+ \frac{1}{2} ( \tens{v}{kl}{pq} \tens{t}{ab}{ij} \density{k}{i} \density{l}{j} - \tens{v}{ux}{pq} \tens{t}{ab}{vy} \cdensity{u}{v} \cdensity{x}{y} )$ \\

12 & ${\cal P}(p,b) {\cal P}(q,j)\, \tens{C}{pb}{qj}$ & $+ \tens{v}{kp}{aq} \tens{t}{ab}{ij} \density{k}{i} - \tens{v}{ip}{yq} \tens{t}{xb}{ij} \density{x}{y}$ \\

\hline

\hline
\end{tabular*}
\end{table*}

Table~\ref{tab:expression} reports all terms resulting from $\hat{C} (s) = [\hat{H}, \hat{T}(s)]_{1,2}$, expressed in terms of the one-particle density matrix ($\density{}{}$), the one-hole density matrix ($\cdensity{}{}$), and density cumulants ($\cumulant{}{}$) of $\mref$.\cite{Mukherjee1997p432,Kutzelnigg:2010iu,Hanauer2012p50}
For convenience, we adopt the Einstein summation convention, and drop the symbol ``$(s)$'' from the cluster amplitudes.
Line 1 corresponds to the fully contracted contribution, while lines 2--5 and 6--12 report the one- and two-body contributions of $\hat{C} (s)$, respectively.
In lines 6--9 and 12, we introduce the index permutation operator ${\cal P}(p,q)$ defined as ${\cal P}(p,q) = 1 - (p \leftrightarrow q)$ to indicate contributions to permutation of the tensor $\tens{C}{pq}{rs}$. 
For example, line 6 should be interpreted as:
\begin{align}
    \tens{C}{pb}{ij} \leftarrow &+ \sum_{a} \tens{t}{ab}{ij} \tens{f}{p}{a}, \\
    \tens{C}{bp}{ij} \leftarrow &- \sum_{a} \tens{t}{ab}{ij} \tens{f}{p}{a}.
\end{align}

\bibliography{LDSRG2-bib}

\begin{thebibliography}{125}
\bibitem [{2()}]{2}%
  \BibitemOpen
  \href@noop {} {}\bibinfo {note} {Due to the high cost of FCI computations,
  the FCI PEC was generated on the grid of 0.1 \AA , while curves for other
  methods were constructed using a much finer grid (0.01 \AA ). To evaluate the
  error with respect to FCI on the finer grid, we perform a fourth-order least
  squares polynomial fit of the MRCISD+Q and FCI energy difference, $f(r) =
  E_\text {MRCISD+Q} (r) - E_\text {FCI} (r) = \sum _{n=0}^{4} a_{n} r^{n}$.
  Thus at a certain atomic distance $r$, the energy deviation for the method
  $X$ relative to FCI is calculated as $\Delta E(r) = E_X (r) - E_\text
  {MRCISD+Q} (r) + f(r)$.}\BibitemShut {Stop}%
\end{thebibliography}
\end{document}